\documentclass[twocolumn]{aastex63}

\newcommand{\comment}[1]{}
\usepackage{lineno}

\shortauthors{Subrayan et al.}

\graphicspath{{./}{figures/}}
\begin{document}
\title{Inferencing Progenitor and Explosion Properties of Evolving Core-collapse Supernovae from \\ Zwicky Transient Facility  Light Curves}

\author[0000-0001-8073-8731]{Bhagya M.\ Subrayan}
\affiliation{Purdue University, Department of Physics and Astronomy, 525 Northwestern Ave, West Lafayette, IN 47907 }

\author[0000-0002-0763-3885]{Dan Milisavljevic}
\affiliation{Purdue University, Department of Physics and Astronomy, 525 Northwestern Ave, West Lafayette, IN 47907 }
\affiliation{Integrative Data Science Initiative, Purdue University, West Lafayette, IN 47907, USA}

\author[0000-0003-1169-1954]{Takashi J. Moriya}
\affiliation{National Astronomical Observatory of Japan, National Institutes of Natural Sciences, 2-21-1 Osawa, Mitaka, Tokyo 181-8588, Japan}
\affiliation{School of Physics and Astronomy, Faculty of Science, Monash University, Clayton, Victoria 3800, Australia}

\author[0000-0002-8360-0831]{Kathryn E.\ Weil}
\affiliation{Purdue University, Department of Physics and Astronomy, 525 Northwestern Ave, West Lafayette, IN 47907 }

\author[0000-0001-9314-0683]{Geoffery Lentner}
\affiliation{Purdue University, Department of Physics and Astronomy, 525 Northwestern Ave, West Lafayette, IN 47907 }

\author{Mark Linvill}
\affiliation{Purdue University, Department of Physics and Astronomy, 525 Northwestern Ave, West Lafayette, IN 47907 }

\author[0000-0003-0776-8859]{John Banovetz}
\affiliation{Purdue University, Department of Physics and Astronomy, 525 Northwestern Ave, West Lafayette, IN 47907 }

\author[0000-0001-6922-8319]{Braden Garretson}
\affiliation{Purdue University, Department of Physics and Astronomy, 525 Northwestern Ave, West Lafayette, IN 47907 }

\author[0000-0002-1521-0479]{Jack Reynolds}
\affiliation{Purdue University, Department of Physics and Astronomy, 525 Northwestern Ave, West Lafayette, IN 47907 }

\author{Niharika Sravan}
\affiliation{California Institute of Technology, Pasadena, USA }

\author[0000-0002-7706-5668]{Ryan Chornock}
\affiliation{Department of Astronomy, University of California, Berkeley, CA 94720-3411, USA}

\author[0000-0003-4768-7586]{Raffaella Margutti}
\affiliation{Department of Astronomy, University of California, Berkeley, CA 94720-3411, USA}

\begin{abstract}

We analyze a sample of 45 Type II supernovae from the Zwicky Transient Facility (ZTF) public survey using a grid of hydrodynamical models in order to assess whether theoretically-driven forecasts can intelligently guide follow up observations supporting all-sky survey alert streams. We estimate several progenitor properties and explosion physics parameters including zero-age-main-sequence (ZAMS) mass, mass-loss rate, kinetic energy, $^{56}$Ni mass synthesized, host extinction, and the time of explosion.  Using complete light curves we obtain confident characterizations for 34 events in our sample, with the inferences of the remaining 11 events limited either by poorly constraining data or the boundaries of our model grid. We also simulate real-time characterization of alert stream data by comparing our model grid to various stages of incomplete light curves ($\Delta t <$ 25 days, $\Delta t < 50$ days, all data), and find that some parameters are more reliable indicators of true values at early epochs than others. Specifically, ZAMS mass, time of explosion, steepness parameter $\beta$, and host extinction are reasonably constrained with incomplete light curve data, whereas mass-loss rate, kinetic energy and $^{56}$Ni mass estimates generally require complete light curves spanning $> 100$ days. We conclude that real-time modeling of transients, supported by multi-band synthetic light curves tailored to survey passbands, can be used as a powerful tool to identify critical epochs of follow up observations. Our findings are relevant to identify, prioritize, and coordinate efficient follow up of transients discovered by Vera C. Rubin Observatory. 

\end{abstract}


\section{Introduction} \label{sec:intro}

The upcoming Legacy Survey of Space and Time (LSST) to be conducted by the Vera C. Rubin Observatory is highly anticipated to revolutionize time domain astronomy \citep{Abell2009LSST}. Its sensitivity ($\sim$ 24 mag), six broadband filters ($u$-$g$-$r$-$i$-$z$-$y$), regular southern-sky patrolling (cadences anticipated between hourly and every few days; \citealt{Marshall2017LSST}), and prompt reporting of transient activity (latency of $\approx 60$ seconds from exposure read-out to alert distribution) will provide opportunities to discover and investigate millions of supernovae (SNe) over its planned ten year lifetime \citep{Ivezic2019}. 

However, managing the massive data sets associated with LSST will be demanding. It will produce $\sim$ 20 TB of raw images every single night, which will be processed rapidly via template subtraction to send out real-time alerts of residual source variability (approximately ten million alerts nightly). Moreover, because LSST photometry alone will generally be insufficient to adequately investigate the transients it will discover \citep{Alves2022}, the survey's success will be partially dependent on other telescopes for supporting observations (see, e.g., \citealt{Najita2016}), with electromagnetic and multi-messenger facilities \citep{Huerta2019}. 

The LSST Corporation (LSSTC) architects along with emerging alert stream data brokers are developing the processes, cyberinfrastructure, and software needed to confront this challenge and help manage upcoming LSST discoveries \citep{Borne2008,Narayan2018}. Data brokers utilize transient classification methods that often employ machine learning \citep{Moller2020,Forster2021,Sooknunan21,Garcia2022}. Current data brokers include ALeRCE\footnote{Automatic Learning for the Rapid Classification of Events, \url{http://alerce.science/}} \citep{Sanchez2021}, ANTARES\footnote{Arizona--NOIRLab Temporal Analysis and Response to Events System, \url{https://antares.noirlab.edu/}} \citep{Matheson2021},  Lasair\footnote{\url{https://lasair.roe.ac.uk/}} \citep{Smith2019}, MARS\footnote{Make Alerts Really Simple, \url{https://mars.lco.global/}}, and FINK\footnote{\url{https://fink-broker.org/}} \citep{Moller2021}, Babamul and PITT-Google. These data brokers are taking on different responsibilities to promptly process, value-add, cross-reference, and classify survey alert streams, which in turn permits users to filter and prioritize targets. 

Working downstream of these data brokers are additional services to coordinate follow up observations, including Target and Observation Managers (TOMs) that permit observers to sort through broker alert streams to plan and trigger follow up \citep{Street2018}.  TOMs have some level of automation, but generally rely largely on humans to make decisions about target prioritization and coordination.  The Recommender Engine For Intelligent Transient Tracking (REFITT; \citealt{Sravan2020}) is an attempt to completely automate transient follow up as an Object Recommender for Augmentation and Coordinating Liaison Engine (ORACLE). REFITT uses data ingested from surveys to predict the light curve evolution of transients, prioritize events based on confidence in its prediction, and finally makes recommendations to observers on targets that need follow up, specific to their observing facility and coordinated among all observing agents. 

To date, decisions about which transients to prioritize for follow up observations with supporting facilities are generally \textit{data-driven}; i.e., based on comparisons to data sets of previously observed events. In this paper, we explore the feasibility of guiding follow-up of core--collapse supernovae (CCSNe) using parameters of \textit{theoretically-driven} forecasts. Our expectation is that prioritizing the underlying physics of transients will make it possible to 1) rapidly recognize transients of desired physical parameter spaces, and 2) identify information-rich epochs in transient evolution for efficient follow-up with limited facilities. 

To this end, in this paper we characterize a sample of 45 light curves of Type II supernovae using publicly available data from the Zwicky Transient Facility (ZTF) \citep{Bellm2019} survey with a grid of theoretical hydrodynamical models spanning various progenitor properties (zero-age-main-sequence (ZAMS) mass, mass-loss history, etc.) and explosion physics (e.g., kinetic energy, $^{56}$Ni synthesized). We compare results between model fits with both complete and incomplete light curves in order to assess whether theoretically-driven forecasts can intelligently guide follow up observations supporting all-sky survey alert streams. ZTF data is used because, among currently operating all-sky surveys that include ATLAS\footnote{Asteroid Terrestrial-impact Last Alert System} \citep{Tonry2018} and ASAS--SN\footnote{All-Sky Automated Survey for Supernovae} \citep{Shappee2014}, ZTF's functioning alert stream best mimics LSST data flow, but at a more manageable scale \citep{Masci2019}. Given that ZTF's cadence, depth, and filters differ from those of LSST, it still serves as an excellent testing ground for developing the infrastructure and software that LSST will require as it starts operating.

The paper is organized as follows: Section \ref{sec:data} describes the treatment of ZTF data in multiple passbands with forced photometry and extinction. In Section \ref{sec:models}, we describe our hydrodynamical models with circumstellar material (CSM) structure constructed using the stellar evolutionary code KEPLER and the radiative transfer code STELLA. Section \ref{sec:parainference} outlines the results from our fitting method and trends in the parameters derived from the fits, and Section \ref{sec:real-time} describes our real-time fitting analysis for ZTF events and an assessment of how model parameters evolve as a function of time. The implications and utility of this work in the backdrop of current and upcoming all--sky surveys including LSST are discussed in Section \ref{sec:discussions}.

\section{Survey Data from ZTF} \label{sec:data}

We used data from the public ZTF alert stream, available as photometry in the ztf--$g$ and ztf--$r$ passbands. We selected 45 ZTF events from \citet{Garretson2021} that are spectroscopically classified as Type II or IIP (Table \ref{tab:events}). For the events that were classified as Type II, we confirmed that the light curves clearly demonstrated some portion of the plateau phase through visual inspection, differentiating them from Type IIL supernovae. The events span across the first four years of ZTF survey starting from 2018--2021. All events in this sample have a minimum of 5 detections, as defined below,  in each ztf--$g$ and ztf--$r$ passbands. We treat this as a representative test sample as the light curves represent a variety of phases in evolution, reasonably span expected redshifts of the survey ($0.010 < z < 0.055$),  and span peak apparent magnitudes approximately between 18--20 mag.

\startlongtable
\begin{deluxetable*}{lclrrccc}
\linespread{1.1}
\tablecaption{ZTF events in this paper \label{tab:events}}
\tablecolumns{8}
\tablewidth{0pt}
\tablehead{
\colhead{ZTF ID} & 
\colhead{TNS} & 
\colhead{TNS}&
\colhead{RA} &
\colhead{Dec} & 
\colhead{Redshift}&
\colhead{No. $g$-band} &
\colhead{No. $r$-band} \\
\colhead{} & \colhead{Classification}&\colhead{Name} & \colhead{(J2000)} & \colhead{(J2000)}& \colhead{($z$)}&\colhead{detections} &\colhead{detections}  
}

\startdata
ZTF18abcpmwh\tablenotemark{$*$}&IIP&SN 2018cur&12:59:09.12&+37:19:00.19&0.015&8(12)&8(40)\\
ZTF18acbvhit&II&SN 2018hle&3:39:28.11&$-$13:07:02.50&0.014&16&15\\
ZTF18acbwasc\tablenotemark{$\dagger$}&IIP&SN 2018hfc&11:01:58.61&+45:13:39.26&0.020&51&65\\
ZTF18acrtvmm\tablenotemark{$*$}&II&SN 2018jfp&3:17:56.27&$-$0:10:10.82&0.023&17&20(2)\\
ZTF18acuqskr&II&SN 2018jrb&8:09:33.69&+15:31:10.55&0.045&7&10\\
ZTF19aakiyed&II&SN 2019awk&15:07:02.58&+61:13:42.37&0.044&17&23\\
ZTF19aaqdkrm&II&SN 2019dod&13:25:49.97&+34:29:43.58&0.034&27&30\\
ZTF19aauqwna&IIP&SN 2019fem&19:44:46.13&+44:42:49.13&0.041&20&47\\
ZTF19aavbkly&IIP&SN 2019fmv&12:29:33.80&+35:46:12.15&0.041&31&29\\
ZTF19aavhblr&II&SN 2019fuo&15:31:43.38&+16:42:49.30&0.050&19&27\\
ZTF19aavkptg&IIP&SN 2019gqs&11:36:15.78&+49:09:12.55&0.038&30&22\\
ZTF19abguqsi&II&SN 2019lsh&22:52:58.52&+0:26:50.53&0.052&14&20\\
ZTF19abhduuo\tablenotemark{$\dagger$}&II&SN 2019lre&1:58:50.79&$-$9:35:05.65&0.018&14&16\\
ZTF19abiahko&IIP&SN 2019lsj&19:36:59.13&$-$11:57:13.63&0.023&8&11\\
ZTF19abqyouo\tablenotemark{$\dagger$}&IIP&SN 2019pbk&7:46:23.89&+64:13:23.79&0.045&11&13\\
ZTF19abvbrve&IIP&SN 2019puv&19:12:36.67&$-$19:25:01.67&0.020&14&42\\
ZTF19acbvisk\tablenotemark{$*$}&II&SN 2019rms&9:00:45.39&+19:44:42.32&0.037&19&37(1)\\
ZTF19ackjvtl\tablenotemark{$\dagger$}&IIP&SN 2019uwd&13:16:20.75&+30:40:48.72&0.019&53&99\\
ZTF19acmwfli\tablenotemark{$\dagger$}&II&SN 2019tza&13:14:04.68&+59:15:04.91&0.028&39&59\\
ZTF19acszmgx&II&SN 2019vew&5:27:49.43&$-$5:21:39.88&0.042&22&22\\
ZTF20aahqbun&II&SN 2020alg&14:27:10.55&+35:55:20.22&0.028&26&41\\
ZTF20aamlmec&II&SN 2020chv&7:46:27.78&+1:57:34.73&0.034&8&9\\
ZTF20aamxuwl&II&SN 2020ckv&11:03:26.48&$-$1:32:27.04&0.037&13&12\\
ZTF20aatqgeo&II&SN 2020fcx&13:40:10.01&+23:20:29.56&0.032&25&16\\
ZTF20aatqidk&II&SN 2020fbj&12:47:47.03&+22:17:10.44&0.034&26&29\\
ZTF20aaullwz&II&SN 2020fch&11:33:24.03&$-$9:20:55.94&0.027&12&10\\
ZTF20aausahr&II&SN 2020hgm&8:31:22.09&+49:13:35.43&0.043&9&10\\
ZTF20aazcnrv&II&SN 2020jjj&14:31:19.63&$-$25:39:31.04&0.023&12&13\\
ZTF20aazpphd&II&SN 2020jww&16:10:51.58&+27:09:42.02&0.046&36&43\\
ZTF20abekbzp&II&SN 2020meu&15:34:44.68&+6:38:53.35&0.041&9&11\\
ZTF20abuqali&II&SN 2020rht&2:30:17.30&+28:36:02.64&0.040&13&12\\
ZTF20abwdaeo\tablenotemark{$\dagger$}&II&SN 2020rvn&21:06:36.34&+17:59:34.86&0.021&28&29\\
ZTF20abyosmd\tablenotemark{$\dagger$}&II&SN 2020toc&8:28:30.14&+17:28:08.52&0.021&20&26\\
ZTF20acjqksf\tablenotemark{$\dagger$}&IIP&SN 2020tfb&6:08:52.14&$-$26:24:46.02&0.048&17&18\\
ZTF20acnvtxy\tablenotemark{$\dagger$}&IIP&SN 2020zkx&11:18:31.84&+6:44:28.84&0.030&10&13\\
ZTF20acptgfl&IIP&SN 2020zjk&5:22:00.02&$-$7:11:20.81&0.037&22&27\\
ZTF21aabygea\tablenotemark{$*$}&II&SN 2021os&12:02:54.08&+5:36:53.15&0.019&31&49(2)\\
ZTF21aaevrjl&II&SN 2021arg&4:31:18.78&$-$10:23:46.95&0.031&14&15\\
ZTF21aafkktu&II&SN 2021avg&11:39:59.00&+14:31:40.65&0.031&31&32\\
ZTF21aafkwtk&II&SN 2021apg&13:41:19.24&+24:29:43.88&0.027&26&37\\
ZTF21aagtqpn\tablenotemark{$\dagger$}\tablenotemark{$*$}&II&SN 2021bkq&18:20:34.83&+40:56:36.28&0.036&33(4)&44(6)\\
ZTF21aaigdly\tablenotemark{$\dagger$}&II&SN 2021cdw&14:05:31.80&$-$25:21:54.51&0.040&17&28\\
ZTF21aaluqkp&II&SN 2021dhx&11:05:10.38&$-$15:21:10.13&0.025&34&25\\
ZTF21aamzuxi&II&SN 2021dvl&7:49:56.10&+71:15:42.11&0.034&27&29\\
ZTF21acchbmn&II&SN 2021zaa&23:46:41.12&+26:44:45.11&0.032&22&19
\enddata
\tablenotetext{$*$}{The reported number of measurements in ztf--$g$ and ztf--$r$  obtained within 150 days of first detection, after running forced photometry for each ZTF event. Additional detections beyond 150 days not included in the analysis are quoted in parenthesis for each band.}
\tablenotetext{\dagger}{No upper--limit constraints before first detection are available for informed priors on explosion date.}
\end{deluxetable*}

We utilize point spread function (PSF)--fit photometry measurements using difference imaging as provided by the ZTF forced-photometry service\citep{IRSA539}. Complete light curves were constructed from the differential flux measurements (\textit{forcediffimflux} and \textit{forcediffimfluxunc})  that the service returns in each band along with upper limits. A signal-to-noise threshold (SNT) $= 3$ and a signal-to-noise ratio (SNU)  $= 5$ were used to declare the measurements as a detection versus an upper-limit \citep{Masci2019}. The photometric zero-points in each passband were used to calculate the differential magnitudes. 

We use the redshift reported by the Transient Name Server (TNS) \citep{GalYam2021TNS} to calculate the distance modulus using \texttt{astropy} \citep{Robitaille2013astropy,Price-Whelan2018astropy} for every ZTF event assuming standard flat $\Lambda$CDM cosmology model with $H_{0} = 70 \: \text{km} \: \: \text{Mpc}^{-1}\: \text{s}^{-1}$ and $\Omega_{0} = 0.3$. The measurements are corrected for Milky Way extinction using dust maps as prescribed by \citealt{Schlegal1998} for each passband, assuming a $R_{V} = 3.1$. In our analysis, for each ZTF event, we only use measurements up to 150 days in the rest-frame from the first detection. We do not account for cosmological K-corrections in this work since we use a sample of low-redshift events. However, because LSST is expected to discover events at higher redshifts, K-corrections for a similar analysis of LSST events would be non-negligible.

\section{Model Fitting} \label{sec:models}

\subsection{Model grid using STELLA}

The hydrodynamic models used in this work are specific to Type IIP, constructed using the multi-group radiation hydrodynamics code STELLA \citep{Blinnikov1998,Blinnikov2000,Blinnikov2006,Moriya2017,Moriya2018, Ricks2019}. In this work, our models have the following parameters: ZAMS mass, kinetic energy of explosion ($E_{k}$), mass-loss rate ($\dot{M}$), steepness of velocity law ($\beta$) associated with the stellar wind, and $^{56}$Ni mass synthesized. The model parameters along with their corresponding values in the grid are described in Table \ref{tab:tab2}. Red supergiant (RSG) pre-supernova progenitors from \citet{Sukhbold2016} were used, which were calculated using the KEPLER code \citep{Weaver1978}, with physics previously discussed (e.g., \citealt{Woosley2002}). A neutron star remnant mass of 1.4 $\rm{M}_\odot$ is assumed and the SN explosions are triggered by putting thermal energy above the mass cut. $^{56}$Ni is assumed to be uniformly mixed up to half of the hydrogen-rich envelope in the mass coordinate.

\begin{deluxetable*}{ccccc}[tp]
\linespread{1.0}
\tablecaption{Prior Distribution on physical parameters used in our sampling method along with values of parameters in our hydrodynamical model grid. \label{tab:tab2}}
\tablecolumns{4}
\tablewidth{0pt}
\tablehead{
\colhead{Parameter} & 
\colhead{Hydrodynamical Model Values} & 
\colhead{In steps}& 
\colhead{Prior Distribution} &
\colhead{Units}
}
\startdata
 \hline
 $t_{\text{exp}}$  & -- & -- &$ U (0,t_{\text{upper-limit}})$ & day \\ 
 ZAMS & 12 -- 16 & 2 &$N (14,3) \in (12,16)$ & $M_\odot$ \\ 
 $E_{k}$ & 0.5 -- 5.0 & 0.5 &$N (1,1) \in (0.5,5) $ & $10^{51}$ erg \\
 $^{56}$Ni   & 0.01 -- 0.1 & 0.01 (0.001,0.2,0.3)  &$N (0.05,0.01) \in (0.001,0.3)$ & $M_\odot$ \\
 $\beta$ & 1 -- 5 &1 &$N (3,2) \in (1, 5)$ & -- \\
 $ - \text{log}_{10}\dot{M}$ & 1 -- 5 & 0.5 &$ U(4,2) \in (5,1)$ & $M_\odot$ $\text{yr}^{-1}$ \\
 $A_{V}$ & -- & -- &$N (\text{ln}(0.05),2) \in (10^{-4}, 2)$ & mag
\enddata
\tablecomments{The ``In steps'' values in parentheses for $^{56}$Ni are the additional values for the parameters present in the model grid.}
\end{deluxetable*}

Our hydrodynamical model grid also incorporates an associated circumstellar material (CSM) density structure attached to these RSG progenitors. The CSM density is given by

\begin{equation}
    \rho_{CSM} (r) = \frac{\dot{M}}{4 \pi v_{wind} r^{2}},
\end{equation}

\noindent where $\dot{M}$ is the mass-loss rate and $v_{wind}$ is the velocity structure associated with the stellar wind. The radial dependency of $v_{wind}$ is given by the velocity law

\begin{equation}
    v_{wind}(r) = v_{0} + \left(v_{\infty} - v_{0}\right)\left( 1 - \frac{R_{0}}{r} \right)^{\beta},
\end{equation}

\noindent where $v_{0}$ is the initial wind velocity with a value $ < 0.01$ km s$^{-1}$, $v_{\infty}$ is the terminal wind velocity $= 10$ km s$^{-1}$, $R_{0}$ is the wind launching radius set to the photosphere of the star, and $\beta$ is the steepness parameter that gives a measure of wind acceleration. RSG progenitors typically have $ \beta > 2 $, owing to slower acceleration of the stellar winds \citep{Mauron2011}. A fixed CSM radius of $10^{15}$ cm is assumed in our models. 

The RSG progenitors with associated CSM density structures were then exploded as thermal bombs with energies ranging from $0.5$ -- $5.0\times10^{51}$ erg using STELLA. The code calculates the resulting light curve of the explosion following the evolution of spectral energy distributions (SEDs) with time at every epoch.  The light curves in various passbands are obtained by convolving the ZTF filter transmission functions to the numerical SEDs. Our final grid of varying progenitor, explosion, and CSM properties is made up of 4206 unique models. \textcolor{black}{A similar model grid used in this work can be found in \citet{Forster2018} for further reference. } The full details of the numerical model grid will be presented in a separate paper (T.J. Moriya, in prep.).

\subsection{Fast Interpolation Method} \label{subsec:fim}

Using Bayesian Inference methods requires the models to be finely sampled within the parameter space. However, because our model grid is neither complete nor uniform, a scale--independent fast interpolation process that can use a non--uniform grid of models was incorporated into our Monte Carlo sampling method following \citet{Forster2018} and \citet{Martinez2020}. This allowed us to quickly interpolate between the models in the grid to sample any combination of values in the parameter space. For a given parameter vector $\vec{\theta}$, the method finds the closest models $\vec{\theta}_{close}$ and weighs them appropriately using 
\begin{equation}
    m(t, \vec{\theta}) = \sum_{\vec{\theta}_{\it{i}} \in  \vec{\theta}_{\textit{close}} }\hat{w}(\vec{\theta},\vec{\theta}_{\it{i}}) m (t, \vec{\theta}_{\it{i}}),
    \label{EQ1}
\end{equation}
where $m(t, t_{\text{exp}}, \vec{\theta})$ is the magnitude for a given $\vec{\theta}$ at time $t$ and the normalized weights are given by

\begin{equation}
    \hat{w} (\vec{\theta},\vec{\theta}_{\it{i}}) = \frac{w(\vec{\theta},\vec{\theta}_{\it{i}})}{\sum_{\vec{\theta}_{\it{j}} \in  \vec{\theta}_{\text{close}}} w (\vec{\theta},\vec{\theta}_{\it{j}}) },
    \label{EQ2}
\end{equation}

where
\begin{equation}
    w(\vec{\theta},\vec{\theta}_{\it{i}}) = \Big( \prod_{\it{j}} \big|\theta^{\it{j}} - \theta_{\it{i}}^{\it{j}} \big| + \delta^{\it{j}} \Big)^{-1} .
    \label{EQ3}
\end{equation}

Equation \ref{EQ3} uses a very small vector $\vec{\delta}$ with the same units as $\vec{\theta}$, which ensures that the weights do not diverge when a given parameter combination exactly matches a model in the grid. This fast interpolation method can be used to calculate light curves for any combination of parameter vector $\theta$ bound by the limits of our model grid.

\subsection{Explosion date and host extinction}

Along with fitting for the parameters of our models, we also fit for the explosion date and host extinction. The time of explosion is calculated as the number of days before the first photometric measurement in each passband.  We defined an informed prior distribution for the time of explosion leveraging the constraints given by the upper limits in each passband for each event. The priors are more constraining if upper limits preceding the first detection are well defined. To calculate the approximate bounds of time of explosion priors, the deepest upper limit available before the first detection is identified.  If no upper-limits are available for an event, we allow a wider distribution for the prior. 

Observed Type II SN light curves can be significantly affected by host extinction \citep{Kasen2009,Matilla2012,Kochanek2012}. We fit for total extinction in the visual band assuming a prior distribution listed in Table \ref{tab:tab2}. We do this by simultaneously fitting ztf--$g$ and ztf--$r$ passbands to infer host extinction. The derived extinction is used to calculate $E(B-V)$ for the host of each event assuming \textit{$R_{V}$} = 3.1.

\subsection{Nested Sampling Methods for Parameter Inference}

We derive posterior distributions of the parameters involved using the python based MIT-licensed Dynamic Nested Sampling package \texttt{dynesty} \citep{Skilling2004} that estimates the Bayesian target distributions. We assigned a combination of uniform and Gaussian distributions as our priors for different parameters. The prior distributions considered in this work are listed in Table \ref{tab:tab2}. The likelihood function incorporates fast interpolation and evaluates how close the observations are with a sample model drawn. Using the above framework, multi-band ZTF observations in ztf--$g$ and ztf--$r$ passbands in their absolute magnitudes are fit to the hydrodynamical models.

Nested sampling methods require all the samples to be identically and independently distributed (i.i.d) random variables drawn from the prior distribution.
We use uniform sampling method widely used for dimensions $< 10$ in the \texttt{dynesty} package \citep{Skilling2006}. In the Bayesian framework all of the inference is contained in the final multi-dimensional posterior, which can be marginalised over each parameter to obtain constraints. The Bayesian evidence is represented by the overall normalisation of this posterior. The nested sampling algorithm converges until the evidence has been estimated to a desired accuracy after accepting or rejecting the samples that are drawn from the prior distribution. 

\startlongtable
\begin{deluxetable*}{lccccccc}
\tablecaption{ Median values from the posterior distribution with 1--$\sigma$ uncertainty when fitted with all data from the events. \label{tab:multi}}
\tablecolumns{8}
\tablehead{
\colhead{ZTF ID} & 
\colhead{ZAMS} & 
\colhead{$E_{k}$} &
\colhead{-log$_{10}\dot{M}$} & 
\colhead{M$_{Ni56}$} &
\colhead{$t_{exp}$}&
\colhead{$\beta$}&
\colhead{$A_{V}$}\\
\colhead{}&
\colhead{($M_{\odot}$)}&
\colhead{ ($10^{51}$ ergs)}& 
\colhead{ ($M_{\odot}\,\text{yr}^{-1}$)}&
\colhead{($M_{\odot}$)}&
\colhead{(days)}&
\colhead{}&
\colhead{(mag)}
}
\startdata
ZTF18abcpmwh&$15.77_{-1.39}^{+0.20}$&$1.67_{-0.29}^{+0.35}$&$4.09_{-0.40}^{+0.38}$&$0.07_{-0.02}^{+0.02}$&$14.81_{-1.58}^{+1.88}$&$2.56_{-0.27}^{+1.20}$&$0.23_{-0.10}^{+0.11}$\\
ZTF18acbvhit\tablenotemark{$\S$}&$13.35_{-0.91}^{+1.68}$&$0.51_{-0.01}^{+0.02}$&$4.00_{-0.09}^{+0.09}$&$0.04_{-0.01}^{+0.01}$&$25.54_{-4.04}^{+3.93}$&$3.98_{-0.10}^{+0.09}$&$0.31_{-0.09}^{+0.09}$\\
ZTF18acbwasc\tablenotemark{$\S$}&$12.47_{-0.31}^{+0.40}$&$0.52_{-0.01}^{+0.01}$&$4.22_{-0.36}^{+0.35}$&$0.10_{-0.02}^{+0.005}$&$48.48_{-1.79}^{+1.06}$&$3.06_{-0.20}^{+0.17}$&$0.02_{-0.01}^{+0.02}$\\
ZTF18acrtvmm&$12.86_{-0.50}^{+0.80}$&$0.99_{-0.05}^{+0.04}$&$2.43_{-0.25}^{+0.15}$&$0.10_{-0.03}^{+0.04}$&$8.81_{-1.00}^{+0.75}$&$3.64_{-1.22}^{+0.14}$&$0.03_{-0.02}^{+0.03}$\\
ZTF18acuqskr&$13.89_{-0.98}^{+1.02}$&$2.90_{-0.39}^{+0.30}$&$3.91_{-0.45}^{+0.47}$&$0.04_{-0.02}^{+0.02}$&$16.72_{-2.38}^{+2.40}$&$3.01_{-0.45}^{+0.46}$&$0.06_{-0.04}^{+0.07}$\\
ZTF19aakiyed&$14.85_{-1.74}^{+1.04}$&$1.02_{-0.20}^{+0.21}$&$3.75_{-0.54}^{+0.52}$&$0.06_{-0.02}^{+0.02}$&$17.29_{-3.09}^{+3.22}$&$2.93_{-0.53}^{+0.82}$&$0.10_{-0.07}^{+0.10}$\\
ZTF19aaqdkrm&$13.14_{-0.81}^{+2.52}$&$1.08_{-0.28}^{+0.25}$&$1.67_{-0.21}^{+0.25}$&$0.05_{-0.01}^{+0.02}$&$16.68_{-2.88}^{+1.95}$&$3.06_{-1.20}^{+0.89}$&$0.12_{-0.08}^{+0.11}$\\
ZTF19aauqwna&$13.71_{-1.01}^{+1.17}$&$1.19_{-0.31}^{+0.30}$&$1.32_{-0.20}^{+0.26}$&$0.08_{-0.02}^{+0.02}$&$16.29_{-3.80}^{+6.28}$&$3.03_{-0.50}^{+0.55}$&$0.05_{-0.03}^{+0.07}$\\
ZTF19aavbkly&$13.07_{-0.75}^{+2.09}$&$1.06_{-0.21}^{+0.43}$&$2.43_{-0.25}^{+0.18}$&$0.02_{-0.01}^{+0.01}$&$7.27_{-0.82}^{+0.51}$&$3.00_{-0.10}^{+0.10}$&$0.18_{-0.10}^{+0.25}$\\
ZTF19aavhblr&$13.46_{-0.97}^{+1.36}$&$0.86_{-0.19}^{+0.28}$&$1.09_{-0.04}^{+0.41}$&$0.07_{-0.01}^{+0.02}$&$24.13_{-10.53}^{+4.25}$&$2.98_{-0.47}^{+0.49}$&$0.05_{-0.04}^{+0.09}$\\
ZTF19aavkptg&$13.14_{-0.80}^{+1.97}$&$1.06_{-0.26}^{+0.45}$&$2.22_{-0.20}^{+0.30}$&$0.03_{-0.02}^{+0.02}$&$8.21_{-1.40}^{+1.09}$&$2.95_{-0.50}^{+0.71}$&$0.38_{-0.17}^{+0.25}$\\
ZTF19abguqsi&$12.57_{-0.40}^{+1.02}$&$1.24_{-0.45}^{+0.73}$&$1.51_{-0.04}^{+0.06}$&$0.01_{-0.01}^{+0.01}$&$13.71_{-1.34}^{+0.88}$&$3.00_{-0.44}^{+0.45}$&$0.17_{-0.10}^{+0.19}$\\
ZTF19abhduuo\tablenotemark{$\S$}&$12.95_{-0.59}^{+0.85}$&$0.52_{-0.02}^{+0.05}$&$4.00_{-0.08}^{+0.08}$&$0.03_{-0.01}^{+0.01}$&$51.02_{-6.87}^{+5.20}$&$3.97_{-0.36}^{+0.47}$&$0.79_{-0.11}^{+0.17}$\\
ZTF19abiahko&$13.11_{-0.71}^{+0.92}$&$0.75_{-0.21}^{+0.32}$&$3.99_{-0.10}^{+0.11}$&$0.02_{-0.01}^{+0.01}$&$25.15_{-5.27}^{+7.76}$&$3.99_{-0.10}^{+0.10}$&$0.54_{-0.36}^{+0.23}$\\
ZTF19abqyouo\tablenotemark{$\S$}&$12.98_{-0.70}^{+1.36}$&$1.39_{-0.23}^{+0.15}$&$3.95_{-0.70}^{+0.58}$&$0.06_{-0.02}^{+0.02}$&$31.37_{-3.42}^{+3.16}$&$3.00_{-0.84}^{+0.85}$&$0.09_{-0.06}^{+0.09}$\\
ZTF19abvbrve\tablenotemark{$\ddag$}&$12.85_{-0.59}^{+1.59}$&$0.51_{-0.01}^{+0.02}$&$3.56_{-0.25}^{+0.57}$&$0.06_{-0.02}^{+0.02}$&$8.27_{-1.16}^{1.47}$&$2.82_{-0.43}^{+0.89}$&$0.13_{-0.07}^{+0.09}$\\
ZTF19acbvisk&$13.82_{-0.86}^{+0.77}$&$0.54_{-0.03}^{+0.07}$&$1.63_{-0.41}^{+0.41}$&$0.10_{-0.02}^{+0.02}$&$20.02_{-6.41}^{+9.31}$&$3.11_{-0.49}^{+0.48}$&$0.02_{-0.02}^{+0.03}$\\
ZTF19ackjvtl\tablenotemark{$\S$}&$14.43_{-1.26}^{+1.05}$&$0.51_{-0.01}^{+0.02}$&$3.88_{-0.80}^{+0.71}$&$0.06_{-0.00}^{+0.01}$&$57.60_{-1.99}^{+2.22}$&$2.47_{-0.83}^{+1.37}$&$0.40_{-0.07}^{+0.07}$\\
ZTF19acmwfli&$13.34_{-0.97}^{+2.19}$&$1.01_{-0.04}^{+0.06}$&$3.82_{-0.79}^{+0.71}$&$0.04_{-0.01}^{+0.01}$&$24.75_{-4.51}^{+2.12}$&$2.69_{-0.21}^{+1.07}$&$0.21_{-0.08}^{+0.10}$\\
ZTF19acszmgx&$13.21_{-0.86}^{+1.40}$&$1.39_{-0.40}^{+0.42}$&$1.03_{-0.03}^{+0.07}$&$0.08_{-0.02}^{+0.02}$&$19.37_{-0.88}^{+0.45}$&$2.98_{-0.47}^{+0.50}$&$0.43_{-0.36}^{+0.25}$\\
ZTF20aahqbun&$15.39_{-2.27}^{+0.51}$&$0.58_{-0.04}^{+0.05}$&$3.01_{-0.48}^{+1.11}$&$0.06_{-0.01}^{+0.01}$&$27.73_{-2.11}^{+1.51}$&$2.55_{-0.18}^{+1.20}$&$0.17_{-0.07}^{+0.07}$\\
ZTF20aamlmec&$13.53_{-1.01}^{+1.48}$&$1.00_{-0.26}^{+0.38}$&$2.37_{-0.40}^{+0.72}$&$0.03_{-0.02}^{+0.02}$&$22.35_{-3.56}^{+3.76}$&$2.99_{-0.49}^{+0.53}$&$0.25_{-0.16}^{+0.20}$\\
ZTF20aamxuwl&$13.48_{-1.00}^{+1.49}$&$1.03_{-0.28}^{+0.37}$&$1.78_{-0.39}^{+0.46}$&$0.06_{-0.01}^{+0.02}$&$14.41_{-3.18}^{+4.21}$&$3.05_{-0.54}^{+0.57}$&$0.08_{-0.06}^{+0.11}$\\
ZTF20aatqgeo&$12.60_{-0.43}^{+1.67}$&$0.92_{-0.25}^{+0.15}$&$2.02_{-0.06}^{+0.10}$&$0.06_{-0.02}^{+0.02}$&$18.60_{-1.54}^{+0.97}$&$2.76_{-0.40}^{+0.94}$&$0.29_{-0.13}^{+0.09}$\\
ZTF20aatqidk&$13.81_{-1.04}^{+1.21}$&$0.61_{-0.04}^{+0.05}$&$2.67_{-0.26}^{+0.37}$&$0.10_{-0.04}^{+0.04}$&$16.88_{-2.26}^{+2.47}$&$3.52_{-1.11}^{+0.26}$&$0.03_{-0.02}^{+0.04}$\\
ZTF20aaullwz&$14.01_{-0.69}^{+0.73}$&$1.03_{-0.08}^{+0.38}$&$2.73_{-0.28}^{+0.33}$&$0.08_{-0.02}^{+0.01}$&$10.65_{-1.94}^{+1.76}$&$3.02_{-0.35}^{+0.37}$&$0.04_{-0.03}^{+0.10}$\\
ZTF20aausahr&$13.74_{-1.02}^{+1.18}$&$2.38_{-0.34}^{+0.46}$&$3.69_{-1.01}^{+0.74}$&$0.03_{-0.02}^{+0.02}$&$27.38_{-3.29}^{+3.29}$&$3.00_{-0.47}^{+0.48}$&$0.12_{-0.08}^{+0.12}$\\
ZTF20aazcnrv&$14.80_{-1.61}^{+1.02}$&$1.26_{-0.39}^{+0.51}$&$3.78_{-0.52}^{+0.51}$&$0.02_{-0.01}^{+0.01}$&$19.69_{-2.43}^{+2.65}$&$3.94_{-0.27}^{+0.57}$&$0.35_{-0.20}^{+0.21}$\\
ZTF20aazpphd&$13.73_{-0.90}^{+0.82}$&$1.00_{-0.02}^{+0.03}$&$1.86_{-0.15}^{+0.33}$&$0.10_{-0.02}^{+0.03}$&$13.94_{-1.60}^{+1.67}$&$3.08_{-0.55}^{+0.45}$&$0.02_{-0.01}^{+0.03}$\\
ZTF20abekbzp&$13.62_{-1.05}^{+1.38}$&$1.54_{-0.46}^{+0.75}$&$3.19_{-0.78}^{+1.08}$&$0.02_{-0.02}^{+0.02}$&$15.27_{-3.34}^{+2.84}$&$2.99_{-0.49}^{+0.51}$&$0.28_{-0.18}^{+0.27}$\\
ZTF20abuqali&$13.81_{-1.22}^{+1.62}$&$1.01_{-0.15}^{+0.18}$&$2.41_{-0.33}^{+0.31}$&$0.08_{-0.02}^{+0.02}$&$17.38_{-2.96}^{+3.00}$&$3.00_{-0.56}^{+0.75}$&$0.07_{-0.05}^{+0.08}$\\
ZTF20abwdaeo&$15.22_{-1.95}^{+0.71}$&$0.57_{-0.03}^{+0.03}$&$2.55_{-0.27}^{+0.26}$&$0.09_{-0.02}^{+0.01}$&$20.09_{-2.49}^{+2.75}$&$3.50_{-1.08}^{+0.28}$&$0.03_{-0.02}^{+0.03}$\\
ZTF20abyosmd\tablenotemark{$\S$}&$12.72_{-0.51}^{+1.61}$&$0.97_{-0.25}^{+0.08}$&$4.00_{-0.44}^{+0.44}$&$0.06_{-0.01}^{+0.01}$&$47.15_{-3.01}^{+1.89}$&$2.97_{-0.49}^{+0.55}$&$1.03_{-0.11}^{+0.09}$\\
ZTF20acjqksf\tablenotemark{$\S$}&$12.83_{-0.59}^{+1.62}$&$1.59_{-0.45}^{+0.63}$&$4.05_{-0.63}^{+0.58}$&$0.03_{-0.02}^{+0.02}$&$28.34_{-1.99}^{+1.16}$&$2.98_{-0.47}^{+0.49}$&$0.40_{-0.22}^{+0.24}$\\
ZTF20acnvtxy\tablenotemark{$\S$}&$13.32_{-0.89}^{+1.49}$&$1.38_{-0.49}^{+0.59}$&$3.89_{-0.76}^{+0.65}$&$0.03_{-0.02}^{+0.02}$&$54.43_{-5.08}^{+3.58}$&$3.00_{-0.48}^{+0.44}$&$0.98_{-0.26}^{+0.28}$\\
ZTF20acptgfl\tablenotemark{$\ddag$}&$12.94_{-0.41}^{+0.42}$&$0.52_{-0.01}^{+0.02}$&$3.97_{-0.11}^{+0.15}$&$0.05_{-0.01}^{+0.01}$&$19.41_{-2.55}^{+2.76}$&$3.00_{-0.10}^{+0.10}$&$0.08_{-0.05}^{+0.07}$\\
ZTF21aabygea&$15.96_{-0.19}^{+0.03}$&$0.57_{-0.02}^{+0.02}$&$3.76_{-0.14}^{+0.18}$&$0.05_{-0.02}^{+0.02}$&$9.84_{-0.98}^{+0.87}$&$3.75_{-0.05}^{+0.05}$&$0.02_{-0.01}^{+0.02}$\\
ZTF21aaevrjl&$13.72_{-1.05}^{+1.19}$&$1.42_{-0.60}^{+0.76}$&$4.00_{-0.10}^{+0.10}$&$0.02_{-0.01}^{+0.02}$&$14.82_{-2.54}^{+3.13}$&$3.00_{-0.09}^{+0.10}$&$0.81_{-0.32}^{+0.30}$\\
ZTF21aafkktu&$14.35_{-1.29}^{+1.21}$&$0.56_{-0.05}^{+0.09}$&$2.11_{-0.40}^{+0.47}$&$0.09_{-0.01}^{+0.01}$&$8.56_{-2.31}^{+3.54}$&$2.66_{-0.33}^{+1.09}$&$0.04_{-0.03}^{+0.05}$\\
ZTF21aafkwtk\tablenotemark{$\ddag$}&$14.36_{-0.85}^{+1.12}$&$0.51_{-0.00}^{+0.01}$&$2.75_{-0.20}^{+0.35}$&$0.04_{-0.01}^{+0.01}$&$13.52_{-2.11}^{+1.84}$&$2.78_{-0.40}^{+0.74}$&$0.05_{-0.03}^{+0.04}$\\
ZTF21aagtqpn&$12.11_{-0.08}^{+0.17}$&$1.44_{-0.10}^{+0.10}$&$4.04_{-0.45}^{+0.45}$&$0.10_{-0.002}^{+0.002}$&$19.70_{-1.40}^{+1.21}$&$2.50_{-0.30}^{+1.23}$&$0.04_{-0.03}^{+0.04}$\\
ZTF21aaigdly&$14.64_{-1.76}^{+1.24}$&$0.67_{-0.07}^{+0.13}$&$2.79_{-0.26}^{+0.43}$&$0.09_{-0.06}^{+0.07}$&$10.60_{-2.14}^{+2.38}$&$3.71_{-1.24}^{+0.08}$&$0.05_{-0.04}^{+0.06}$\\
ZTF21aaluqkp&$15.14_{-2.06}^{+0.77}$&$0.54_{-0.02}^{+0.04}$&$3.70_{-0.42}^{+0.49}$&$0.09_{-0.03}^{+0.01}$&$16.11_{-2.75}^{+2.74}$&$2.60_{-0.19}^{+1.16}$&$0.07_{-0.04}^{+0.06}$\\
ZTF21aamzuxi&$14.48_{-1.36}^{+1.25}$&$1.49_{-0.09}^{+0.08}$&$3.72_{-0.51}^{+0.49}$&$0.09_{-0.06}^{+0.05}$&$7.85_{-1.31}^{+1.41}$&$2.56_{-0.23}^{+1.20}$&$0.05_{-0.04}^{+0.05}$\\
ZTF21acchbmn&$14.14_{-1.35}^{+1.62}$&$1.51_{-0.08}^{+0.13}$&$2.17_{-0.24}^{+0.32}$&$0.09_{-0.02}^{+0.01}$&$8.96_{-0.89}^{+0.66}$&$3.32_{-0.92}^{+0.46}$&$0.05_{-0.03}^{+0.05}$
\enddata
\tablenotetext{$\ddag$}{Fits to explosion energies are very close to the model grid parameter boundaries.}
\tablenotetext{$\S$}{Events with relatively less confident inferences due to poor data quality (including missing phases of light curve along with no constraints on upper limits)}
\end{deluxetable*}

\begin{figure*}[tp]
	\centering
\includegraphics[width = 0.9\linewidth]{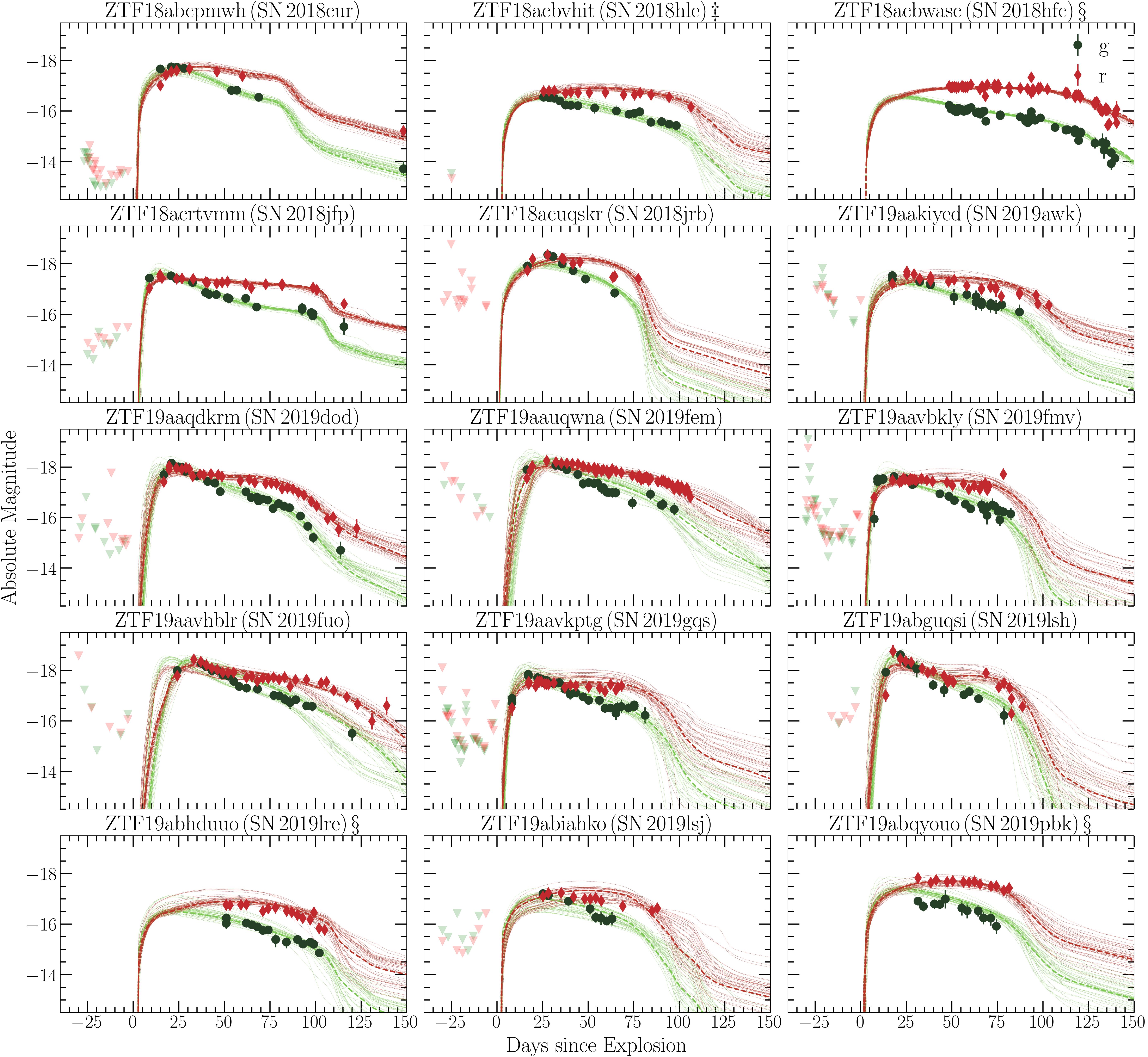}
    \caption{The observed ZTF light curves and our model fits plotted with respect to derived time of explosion . The family of light curves in each panel represents 150 models randomly sampled from the derived posterior probability distribution in individual bands respectively. The dashed curve represents the best fit model for each ZTF event. The upper limits before first detection that are used to constrain explosion date are plotted. The median values and their $1-\sigma$ uncertainity for the parameters are listed in Table \ref{tab:multi}. ZTF events with no upper limits and poorly constraining rise data are flagged with $\S$. ZTF events whose fits approached model boundaries are flagged with $\ddag$.  Additional plots for the remaining ZTF events listed in Table \ref{tab:events} are provided in the Appendix.}
    \label{fig:grid}
\end{figure*}

\section{Parameter Inference and Trends of Complete Light Curves} 
\label{sec:parainference}

Using our Bayesian inference procedure to fit all ZTF events with our model grid, we derived the posterior probability distribution for all parameters of each Type II supernova. Our models were fit confidently to 34 events. The fits for the remaining 11 events were limited by poor upper limits prior to first detection (flagged with $\S$) and the hydrodynamical model grid boundaries (flagged with $\ddag$). Figure \ref{fig:grid} shows the samples of the posterior probability distributions for the first 15 ZTF events. The light curve fits for the remaining events are plotted in an extended version of Figure \ref{fig:grid} in Appendix. \textcolor{black}{The observed data points are shifted with respect to the inferred explosion date and corrected for host extinction derived from the fits. The uncertainity in host extinction and explosion date is not represented in Figure \ref{fig:grid}.} The dashed line in each passband represents the best fit model. Table \ref{tab:multi} lists the inferred median values for the seven parameters with their 16th and 84th percentile confidence regions.
\textcolor{black}{We note that these $1-\sigma$ uncertainities of the parameters only reflect how well the model fits the observed data and do not take into account any uncertainties related to the assumptions used to create the model grid itself.} Representative corner plots of posterior probability distributions are shown in the Appendix.

\begin{figure*}[ht]
	\centering
	\includegraphics[width=\textwidth]{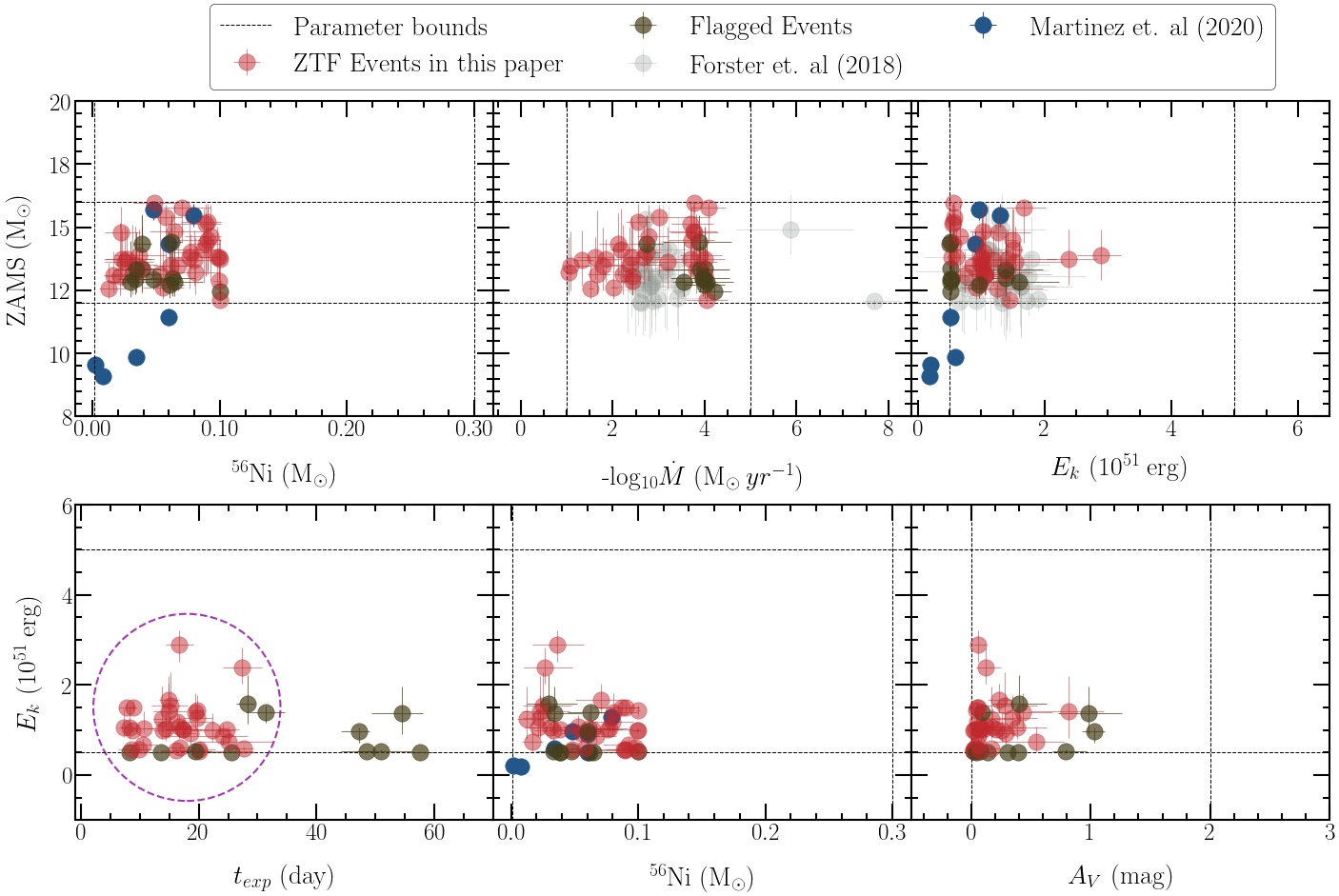}
    \caption{Top panel: ZAMS mass plotted against $^{56}$Ni mass, mass--loss rate, and kinetic energy. The blue and grey circles are values that were found for SNe II in \citealt{Forster2018} and \cite{Martinez2020}. Bottom panel: Kinetic energy plotted against time of explosion with respect to first detection, $^{56}$Ni mass, and host extinction ($A_{V}$). The grey dotted lines represents the parameter bounds of our model grid. The purple circle in the bottom left panel demarcates events that required a narrower prior distribution for the explosion date constrained either from upper limits or rise-time data. The green circles represent flagged ZTF events with low confidence model fits.}
    \label{fig:correlation}
\end{figure*}

ZTF21aabygea (SN 2021os) and ZTF18abcpmwh (SN 2018cur) yielded the highest ZAMS masses ($15.96^{+0.03}_{-0.19}$ M$_{\odot}$ and $15.77^{+0.20}_{-1.39}$ M$_{\odot}$, respectively),  while ZTF21aagtqpn (SN 2021bkq)  has the lowest ($12.11^{+0.17}_{-0.08}$ M$_{\odot}$). The highest explosion energies are seen for ZTF18acuqskr (SN 2018jrb) and ZTF20aausahr (SN 2020hgm) ($2.90^{+0.30}_{-0.39}\times10^{51}$ erg and $2.38^{+0.46}_{-0.38}\times10^{51}$ erg , respectively). Shown in Figure \ref{fig:grid}, these two events correspond to steeper and early declines in the light curves as compared to low energetic events ($\sim 0.5\times10^{51}$ erg ) including ZTF19ackjvtl (SN 2019uwd) and ZTF19acbvisk (SN 2019rms). These observed short-lived plateaus and fast declines are in agreement with previous results for other high energy CCSNe \citep{Valenti2016,Rubin2017,deJaegar2019,Barker2021}. The more energetic events in our sample also show increased peak luminosity in both ztf--$g$ and ztf--$r$ bands \citep{Galbany2016,Valenti2016,Sanders2015}. 

For seven events, our estimates of kinetic energy favor the minimum parameter value of our grid (0.5 $\times10^{51}$ erg).  We flag these events as ones for which our model fits are less confident, since the actual kinetic energy may be significantly lower. 

Most events (37 out of 45) in our sample tend to favor mass--loss rates between $10^{-4.0}-10^{-2.0} \text{M}_{\odot} \text{yr}^{-1}$. However, a non-negligible number of events (8 out of 45) yield higher mass-loss rate estimates $\leq 10^{-2.0}  \text{M}_{\odot} \text{yr}^{-1}$. This finding is consistent with \citet{Forster2018}, who reported higher mass loss rates to be correlated with early and steep rises in the light curves of many Type II SNe from the High cadence Transient Survey (HiTS) \citep{Forster2016}, possibly due to shock-breakout in dense CSM \citep{Moriya2011,Morozova2015,Moriya2017,Morozova2018,Moriya2018,Bruch2021,Haynie2021}. \textcolor{black}{We find that the fits for the steepness parameter $\beta$ are most likely prior dominated and favour values closer to $\beta \sim 3$, consistent with slowly accelerating winds found in red supergiants \citep{Baade1996}.}

ZTF19abguqsi (SN 2019lsh) produced the least $^{56}$Ni mass of $0.01 \pm 0.01 $ $\text{M}_{\odot}$. Other ZTF events in the sample have estimates of $^{56}$Ni mass ranging from 0.02--0.1 $\text{M}_{\odot}$. ZTF19aabvkly (SN 2019fmv), ZTF19abiahko (SN 2019lsj) and ZTF20aazcnrv (SN 2020jjj) have relatively short-lived plateau regions in their light curves and are associated with lower estimates of synthesized $^{56}$Ni mass (see Figure \ref{fig:grid}). Events with higher estimates of $^{56}$Ni mass have long--lived plateaus as compared to the the events with lower $^{56}$Ni mass estimates and faster declines in their light curves. These results are in agreement with previous analyses of SNe Type II that consider how $^{56}$Ni mass  affects light curve evolution \citep{Eastman1994,Melina2013THESIS,Faran2014,Kozyreva2019}.

ZTF20abyosmd (SN 2020toc) has the highest host extinction ($ A_{V} = 1.03^{+0.19}_{-0.11}$ mag), followed by ZTF21aaevrjl (SN 2021arg) ($A_{V}=0.81^{+0.30}_{-0.32}$ mag). ZTF18acbwasc (SN 2018hfc) and ZTF21aabygea (SN 2021os) have negligible host extinction values (both $A_{V}=0.02^{+0.02}_{-0.01}$ mag). All ZTF events in our sample show host extinction values ranging from $ A_{V} = 0.01 - 1.1$ mag, typical for CCSNe hosts \citep{Pastorello2006,Maguire2010,Faran2014}.

\subsection{Correlations between physical parameters}

We plot our inferred values with comparison to other similar works in literature \citep{Forster2018, Martinez2020} in Figure \ref{fig:correlation}. \citet{Forster2018} in their work uses hydrodynamical models on HiTS Type II supernovae that include estimates of circumstellar environment, while \citet{Martinez2020} do not include CSM structure in their models. Figure \ref{fig:correlation} shows the bounds of our model grid with grey dotted lines for each parameter in the plot. The ZTF events whose energy value fits approached the bounds of the model grid are marked in Table \ref{tab:multi}. We represent events that had relatively narrower prior distribution for explosion date inside the purple circle in Figure \ref{fig:correlation}. These events had either constraining upper limits before first detection or enough rise time data to make an educated guess on the upper bound of prior distribution for explosion date. 

\textcolor{black}{Generally, our parameter estimates are less confident for ZTF events with poor data quality, such as those missing phases of light curves combined with unavailable upper limit constraints and whose parameter values approached the parameter boundaries of the model grid (see Table \ref{tab:multi}). These events are represented by green circles in Figure \ref{fig:correlation} as flagged events. Among 7 out of 11, ZTF18acbwasc (SN 2018hfc), ZTF19abhduuo (SN 2019lre), ZTF19abqyouo (SN 2019pbk), ZTF19ackjvtl (SN 2019uwd), ZTF20abyosmd (SN 2020toc), ZTF20acjqksf (SN 2020tfb) and ZTF20acnvtxy (SN 2020zkx) can be grouped together as events that have higher values for time of explosion ($>$ 25 days). This could be attributed to the fact that the prior distribution was too broad as there were no upper limits for these events as provided by the ZTF survey. The remaining 4 ZTF events whose fits approached model boundaries are ZTF18acbvhit (SN 2018hle), ZTF19abvbrve (SN 2019puv), ZTF20acptgfl (SN 2020zjk), and ZTF21aafkwtk (SN 2021apg).}

\begin{figure}[ht]
	\centering
	\includegraphics[width = 1.10\linewidth]{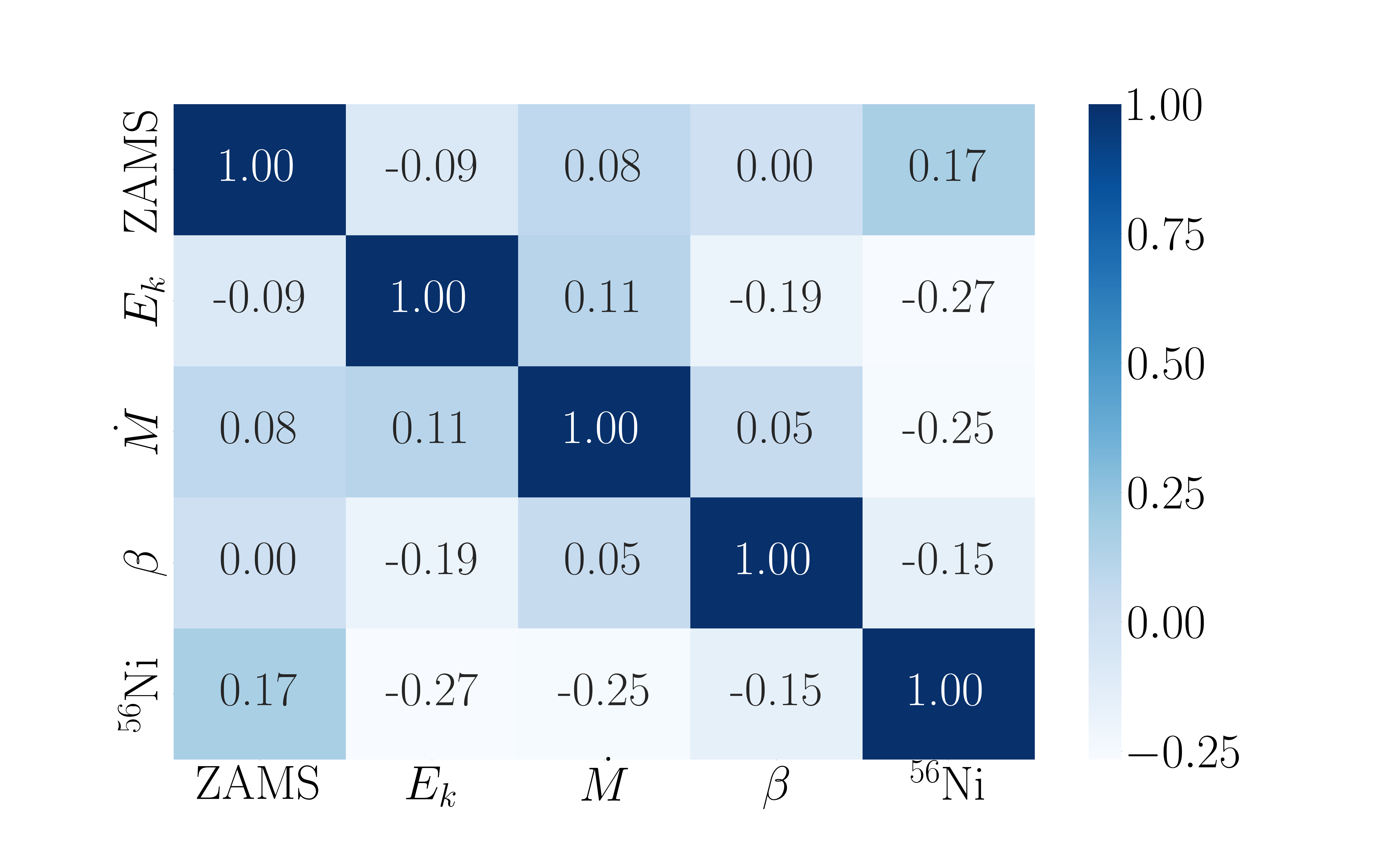}
    \caption{Pearson correlation matrix showing correlation coefficients that are color-coded for different physical parameters in the analysis. No significant correlations between physical parameters were found within the sample of Type II SNe used in this study. }
    \label{fig:correlation2}
\end{figure}

\textcolor{black}{Moreover, we performed a Pearson correlation analysis to check if we can find any correlations between the physical parameters. However, we were unable to find any significant correlations (\textit{r} $> +0.5$ or \textit{r} $< -0.5$) within the sample studied. The correlation matrix is shown in Figure \ref{fig:correlation2}. The modest mass range in our model grid (12 - 16 M$_{\odot} $) limits our ability to make strong claims on any potential ZAMS dependencies. For example, in Figure \ref{fig:correlation}, the correlations in \citet{Martinez2020} becomes only clear when the ZAMS range is extended to 10 M$_{\odot}$. Our hydrodynamical grid explores new parameter spaces that include mass-loss properties and this can possibly introduce additional degeneracies.}

\section{Real-time Parameter Evolution}{\label{sec:multi-epoch}} 
\label{sec:real-time}

We analyzed how each of the parameter fits evolved as a function of time, termed \textit{real-time characterization}. We were motivated to characterize how well or poorly the model parameter fits and their uncertainties at fractional light curve stages anticipated the values found from complete light curves. We compared our model grid to three regimes of incomplete light curves with respect to the first detection: 1) $\Delta t  \leq 25$ days; 2) $\Delta t \leq 50$ days; 3) all available data.  

\begin{figure*}[tp]
	\centering
	\includegraphics[width=0.9\linewidth]{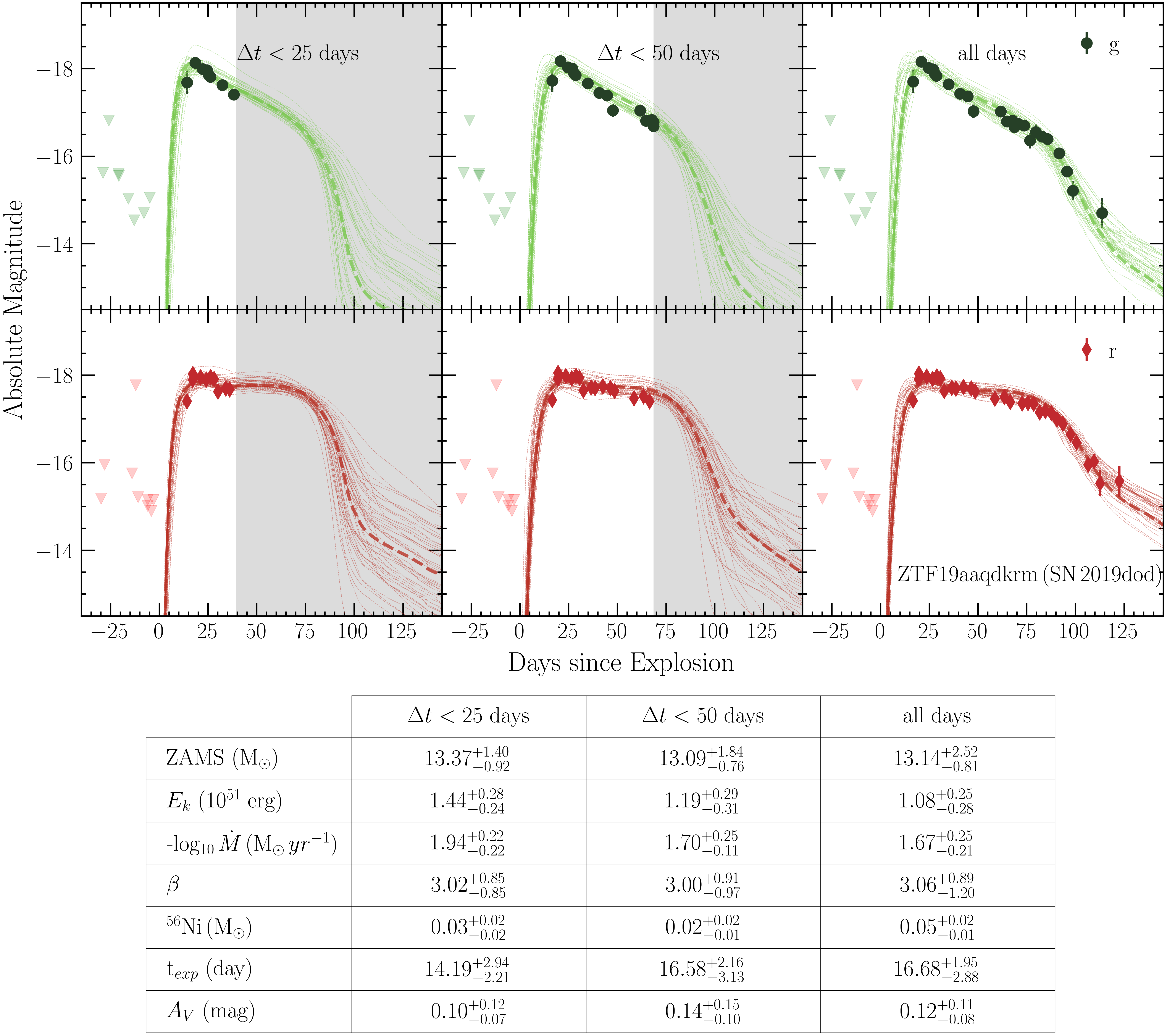}
    \caption{Multi-epoch real-time characterization for ZTF19aaqdkrm (SN 2019dod). The left panel shows model fits to only data within first 25 days of detection in each individual band. The middle panel shows model fits using data within the first 50 days of detection. The right panel uses all  data. The light curves are plotted with respect to the derived time of explosion and host extinction from model fits at each epoch. The table below shows parameter estimates derived from fits at each epoch.}
    \label{fig:dkrm}
\end{figure*}

Figure \ref{fig:dkrm} shows a detailed evolving characterization for the event ZTF19aaqdkrm (SN 2019dod). As expected, the fits and estimates of the parameters change  with time as the supernova evolves and additional measurements are incorporated into the fitting process. Within $\Delta t \leq 25$ days and $\Delta t \leq 50$ days, the fits favor lower $^{56}$Ni masses and higher energies. As data in both bands accumulate, the fits favor higher $^{56}$Ni masses and lower energies. At $\Delta t \geq 50$ days, the portion of the light curve powered by hydrogen recombination of the event starts to fall off, giving better estimates on ZAMS and $^{56}$Ni masses. 

\begin{figure*}[tp]
	\centering
	\includegraphics[width=0.9\textwidth]{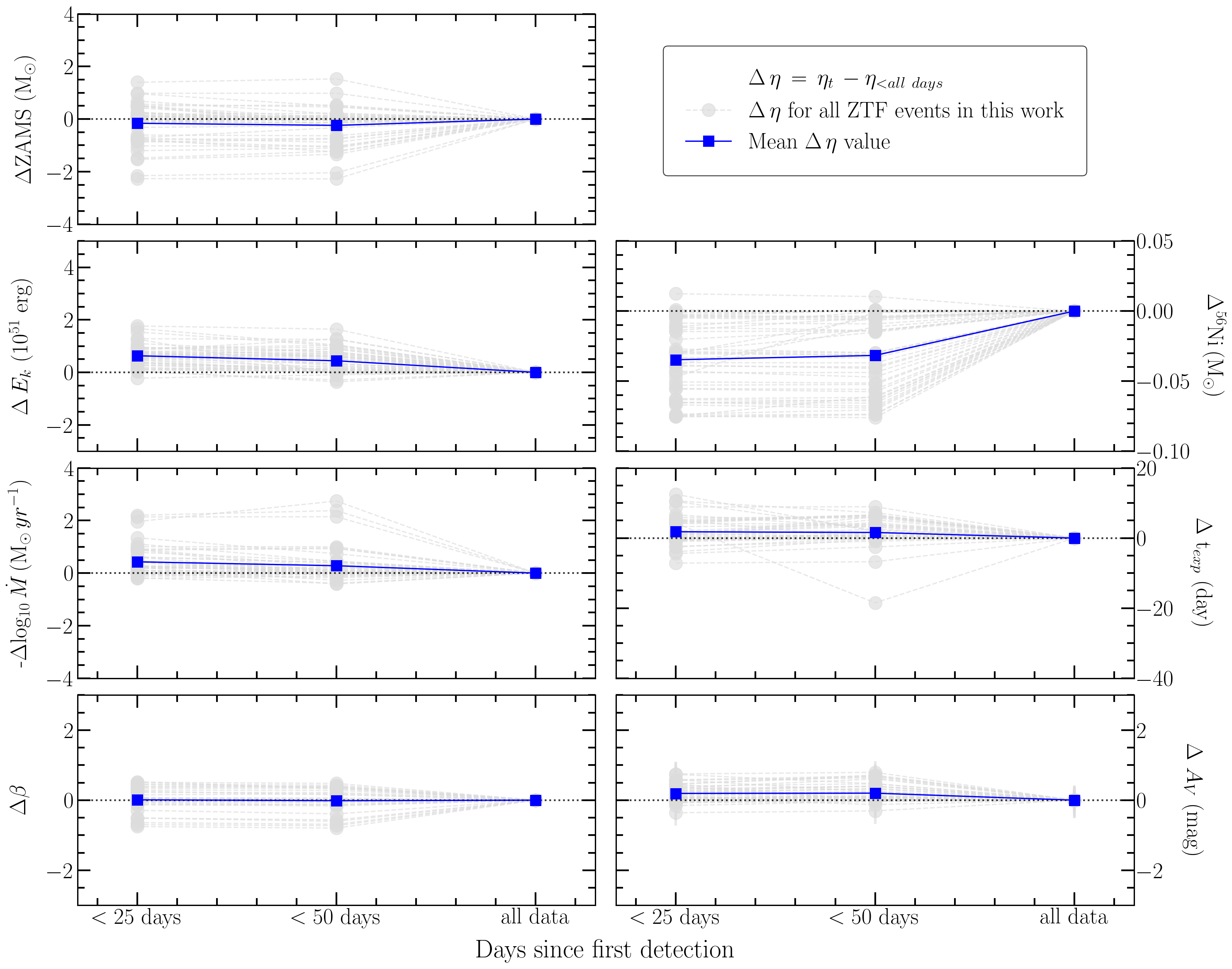}
    \caption{Real-time parameter evolution for all the listed ZTF events. The average change in parameter values (averaged over all 45 ZTF events) at each epoch with respect to the final epoch with all data are represented with the blue squares from top left to bottom right. The order of the parameters are as follows: ZAMS mass, kinetic energy of explosion ($E_{k}$), mass-loss rate ($\dot{M}$), steepness of velocity law ($\beta$) associated with the stellar wind, and $^{56}$Ni mass synthesized, explosion date with respect to first detection, and host extinction ($\text{A}_{V} $).}
    \label{fig:uncertain}
\end{figure*}

We performed this same analysis on all 45 ZTF events, where similar trends are seen. Figure \ref{fig:uncertain} shows the difference in parameter values of our fits with respect to the final epoch with all the data as the event unfolds. Our analysis shows that the explosion energies and mass--loss rates for the events are initially overestimated and tend to favor lower final values when all data is included in the fit. This is opposite to what we see in case of $^{56}$Ni mass, as it is underestimated with only few measurements and consistently favors higher values for many events during later stages of light curve evolution as the recombination drop--off starts to unfold. 

\textcolor{black}{The most confident estimate of $^{56}$Ni mass is inferred when the hydrogen recombination phase ends and the radioactive decay phase starts. This results in epochs with all the data yielding higher estimates of $^{56}$Ni mass. With  kinetic energy, the peak luminosity and decline rate of the light curve plays a significant role in estimating the energetics of the event. As a result, as more data become available at later epochs, more reliable estimates of kinetic energy are seen.} The inferred values of ZAMS, host extinction, explosion date and the $\beta$ remain nearly constant as the light curve evolves. 

\section{Discussion}{\label{sec:discussions}}

\subsection{Parameter Space Degeneracy}
{\label{subsec:degeneracies}}

A major challenge encountered when modelling only two passbands provided by public ZTF light curves was model degeneracy. Specifically, the probability that different combinations of explosion and progenitor parameters can potentially lead to the same light curve posed difficulties in converging to a unique solution in our fitting method. As noted in previous works, the hydrodynamical modelling approach has led to larger estimates of progenitor properties, especially ZAMS estimates, when compared to other approaches like pre-SN explosion imaging \citep{Utrobin2008,Utrobin2009,Maguire2010,Sanders2015}. \textcolor{black}{In the Appendix, posterior distribution of three ZTF events at various epochs are examined using kernel density estimation (KDE) analysis, along with a calculation of the number of modes. In a similar analysis for all 45 events, we find that all the ZTF events have multi-modal posteriors and temporal evolution of modes cannot be generalized.}

\citet{Goldberg2019} recognize the challenges involved in breaking the degeneracies between ejecta mass, explosion energy and progenitor radius, and argue that in order to do so requires an independent measurement of one of the parameters. The scaling relationships used in their work yield families of explosions with varied parameters that can reproduce similar light curves. \citet{Hillier2019} also highlight calculations of similar photospheric phases for well-sampled Type II supernovae in their multi-band and spectroscopic modelling. 

\citet{Martinez2020} attempt to partially lift this degeneracy issue by fitting photospheric velocity information from their models to velocity measurements obtained of SNe during the plateau phase. However, \cite{Goldberg2019} argue that only the ejecta velocities measured during the initial shock cooling phase can be useful to break these degeneracies seen in the parameters. Our analysis focused on photometry only, and future work can investigate whether use of kinematic information from spectroscopy can better constrain parameter selection.

\subsection{Bolometric vs Pre-computed Multi-band Inference}

Our analysis adopts the approach of fitting observations to synthetic multi-band photometry derived from theoretical hydrodynamical models. \textcolor{black}{Similar works like \citet{Nicoll2017} and \citet{Guillochon2018} use semi-analytical, black body spectral energy distribution (SED) models to fit multi-band photometry for transients}. These procedures contrast with typical methods that first construct bolometric light curves from multi-filter and/or multi--wavelength observations, that are in turn compared to model bolometric light curves. Each method has associated uncertainties. In the case of synthetic model photometry, uncertainties  arise from assumptions in opacity treatment at different frequencies in STELLA. In the case of creating bolometric light curves from interpolated observed data sets, the uncertainties stem from potential gaps in photometry cadence, limited passbands, and potentially few data points overall to fit against. 

\textcolor{black}{For real time characterization of events, we found that a proper Bayesian inferencing of explosion parameters for large numbers of supernovae is most efficiently conducted with pre-computed grids of models. Computing models in real-time per event will lead to duplicative efforts and incur computational time costs. With our method, the fitting is more rapid, is more flexible, and the associated uncertainties are less significant.}

\subsection{Intelligent Augmentation}

Our analysis only uses ZTF public data in ztf--$g$ and ztf--$r$ passbands to make inferences about Type II explosion parameters. Our parameter fits could be further constrained with observations at other passbands, and it is worthwhile to consider \textit{which passbands} at \textit{which epochs} are most constraining. To this end, inferencing transients in real time in order to make on-the-fly decisions about optimal follow-up in complementary passbands is needed \citep{Carbone2020,Sravan2021}. 
Generally, most constraining for inferring Type II properties are observations at early and late phases of light curve evolution. At early phases, UV observations best sample shock breakout and circumstellar interaction  \citep{Gezari2015,Ganot2016,Soumagnac2020,Haynie2021,Gallan2022}. At late phases, near- and mid-infrared passbands provide diagnostics that best follow ejecta cooling and dust formation \citep{Szalai2013,Bianco2014,Tinyanont2016}. Optimally augmenting all-sky survey photometry in real-time in this way can enhance opportunities to generate large samples of core-collapse supernovae sufficiently observed to perform population and host-environment studies  \citep{D'Andrea2010,Anderson2014,Sanders2015,Schulze2021}.

\begin{figure*}[tp]
	\centering
	\includegraphics[width=0.9\linewidth]{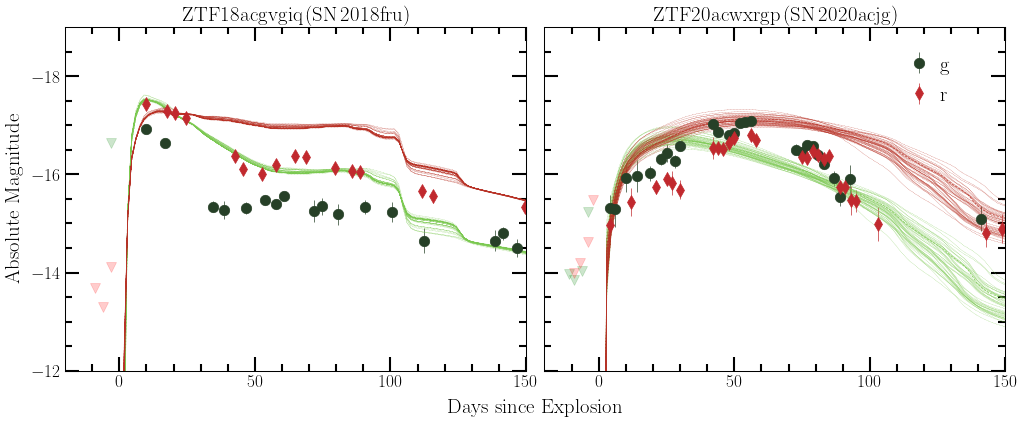}
    \caption{Spectroscopically classified Type II supernovae with anomalous light curves identified in this work. Only the upper limits prior to the first detection used for deriving the fits are plotted.}
    \label{fig:anomalous}
\end{figure*}

\subsection{Anomaly Detection}

Our work uncovered two examples of anomalous Type II supernovae, which are not included in our sample of 45 events.  We present the unusual light curves of ZTF18acgvgiq (SN 2018fru) and ZTF20acwxrgp (SN 2020acjg) in Figure \ref{fig:anomalous}.  The fits for the entire light curves converged to a model solution with very poor likelihood scores resulting in inaccurate inferences. Consequently, we were unable to characterize these events with our current model grid. The anomalous nature of these events could be identified via poor model fits as early as $\Delta t < 25$ days. \textcolor{black}{The cumulative log-evidence (\textit{logZ}) inferred for these two events were above -200 indicating very poor likelihood estimates when comparing the models with the data. For the events with good fits, the \textit{logZ} estimates were under -20. These high \textit{logZ} values indicate that the models were unable to converge to a target distribution of the parameters for the two anomalous events.}

This experience shows that real-time inferencing can be used as a way to identify targets that deviate from normal theoretical predictions. Such real-time analyses for detecting anomalies \citep{Pruzhinskaya2019,Soraisam2020,Villar2020Ano,Ishida2021,Villar2021Ano,Galarza2021} can be automated into an ORACLE such as REFITT to motivate rapid spectroscopic follow--up of non-traditional CCSNe.

\section{Conclusions}\label{sec:conclusion}

In this paper we have characterized 45 Type II supernovae using only products from the public ZTF survey (i.e., in ztf-$g$ and ztf-$r$ passbands) using a grid of theoretical hydrodynamical models. Our grid parameters span multiple supernova progenitor and explosion properties, as well as the time of explosion with respect to the first detection and host extinction. We compare results between complete and fractional light curves to determine which parameters are most robust to incomplete photometric data sets. This effort is to assess whether opportunities exist for theoretically-driven forecasts to inform when follow up observations are needed to support all-sky survey alert streams. The following conclusions are made:

\begin{itemize}

    \item We obtain confident characterizations for 34 SNe II in our sample. Inferences of the remaining 11 events are limited either by poorly constraining data or the boundaries of our model grid. The properties of these well-fitted events broadly follow those reported in previous analyses of SNe II.

    \item In cases where fitted parameters derived from complete vs.\ incomplete data sets are compared, some parameters are more reliably determined at early epochs than others. The explosion energy, host extinction and mass-loss rate parameters are overestimated during initial phases of evolution, while the $^{56}$Ni mass is underestimated. The ZAMS mass and $\beta$ estimates do not change significantly at different phases.
    
    \item  The explosion date is a very sensitive parameter that requires well--constrained pre-explosion upper--limits from the survey for confident inferences.  Generally, we found parameter estimates to be less reliable for ZTF events with poor data quality, such as missing phases of the light curve along with poor upper limit constraints.
    
    \item Real-time Bayesian inferencing of progenitor and explosion parameters for large numbers of CCSNe from all-sky surveys demands a pre-computed grid of models. Creating synthetic model light curves in respective all-sky survey passbands catalyzes real-time characterization of evolving transients by avoiding challenges associated with constructing bolometric light curves with sparse and incomplete photometry.

\end{itemize}

Our work has demonstrated that hydrodynamical model grids for CCSNe along with statistical analyses can provide opportunities to enhance scientific return from all-sky surveys that provide live alert streams. Theoretically-driven predictions can be leveraged to efficiently coordinate worldwide observing facilities to conduct follow up observations that augment survey light curves to optimally achieve scientific objectives \citep{Bianco2014,Modjaz2019,Kennamer2020,Sravan2020,Anand2021}. 

For example, real-time characterization can identify and prioritize transients that fall within certain parameter spaces of interest, including the extreme high and low ends of kinetic energy or $^{56}$Ni mass. Likewise, theoretical forecasts can identify and prioritize follow up photometry at critical phases of transient evolution, including monitoring the plateau drop-off of SNe II light curves that provides information needed to improve estimates of kinetic energy and ZAMS and $^{56}$Ni masses. Ideally, predicting transient evolution using the underlying physics of transients can be incorporated into a TOM or ORACLE that can efficiently recommend targets for follow-up at information-rich epochs \citep{Djorgovski2016,Street2018,Kasliwal2019,Sravan2020,Agayeva2021}.

Our future work relies on an expanded grid of hydrodynamical models exploring larger parameter ranges, including varying degrees of $^{56}$Ni mixing within the inner layers of the progenitors, and information on photospheric velocity that can be used to potentially break degeneracies between parameters. It will also expand synthetic photometry to all six passbands of LSST. Although our work focuses on Type II CCSNe, our methods can be easily applied to identify, prioritize and coordinate follow-up of other transients discovered by Vera C. Rubin Observatory.

\newpage
\section*{Acknowledgements}

The authors would like to thank the anonymous referee for helpful comments that have significantly improved this paper. We also acknowledge helpful discussions with Thomas Matheson, Mariana Orellana, and Melina Bersten. The ZTF forced-photometry service was funded under the Heising-Simons Foundation grant
\#12540303 (PI: Graham). Numerical computations were in part carried out on PC cluster at Center for Computational Astrophysics (CfCA), National Astronomical Observatory of Japan. D.~M.\ acknowledges NSF support from grants PHY-1914448, PHY- 2209451, AST-2037297, and AST-2206532.

\software{ KEPLER \citep{Weaver1978}, STELLA \citep{Blinnikov1998,Blinnikov2000,Blinnikov2006,Moriya2017,Moriya2018, Ricks2019}, astropy\citep{Robitaille2013astropy,Price-Whelan2018astropy}, dynesty \citep{Skilling2004}}

\clearpage

\appendix{\label{appendix:1}}

\begin{figure*}[th!]
	\centering
\includegraphics[width = \textwidth]{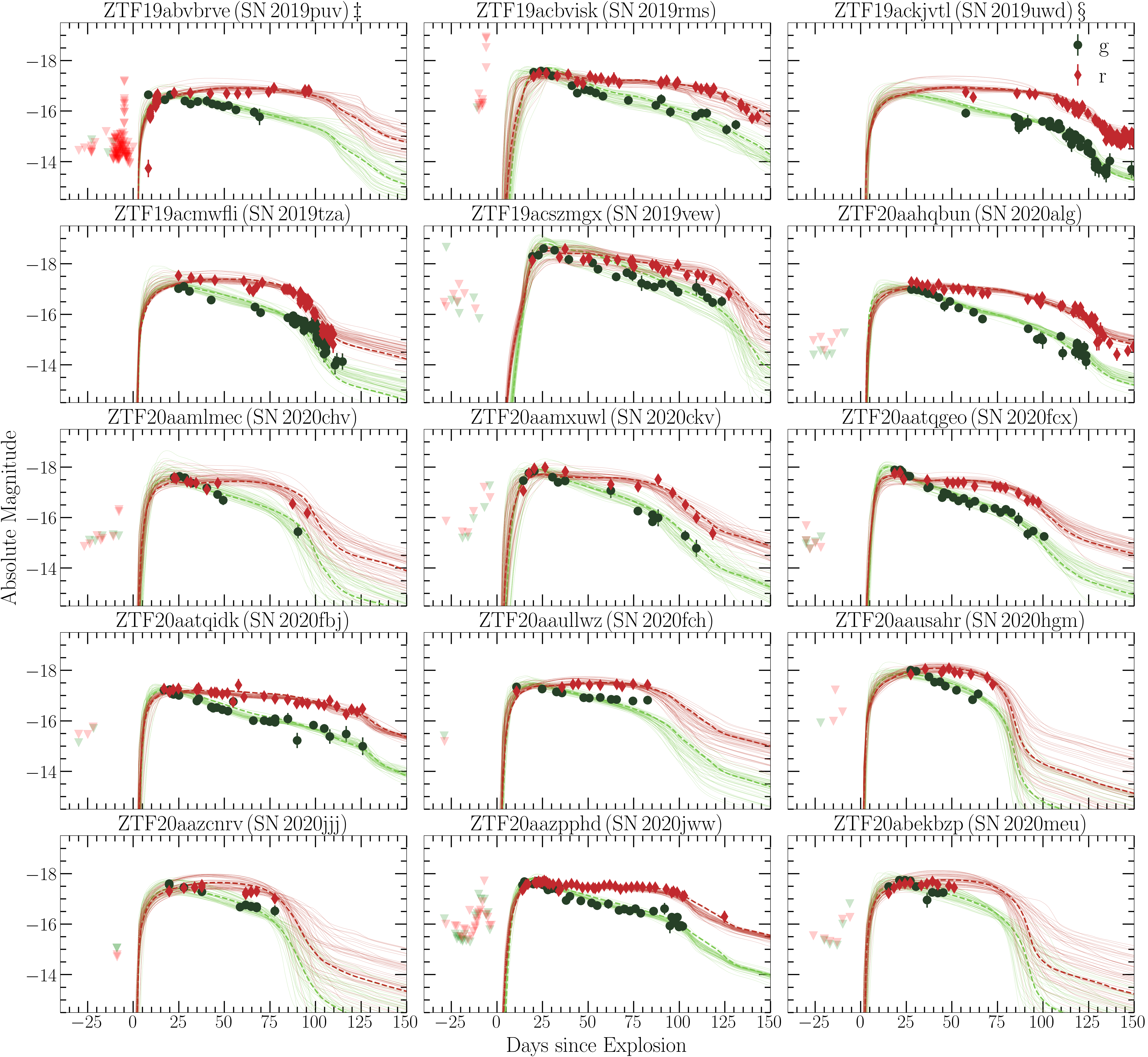}
    \caption{Continued from Figure \ref{fig:grid}. }
    \label{fig:grid_2}
\end{figure*}

\begin{figure*}[th!]
	\centering
\includegraphics[width = \textwidth]{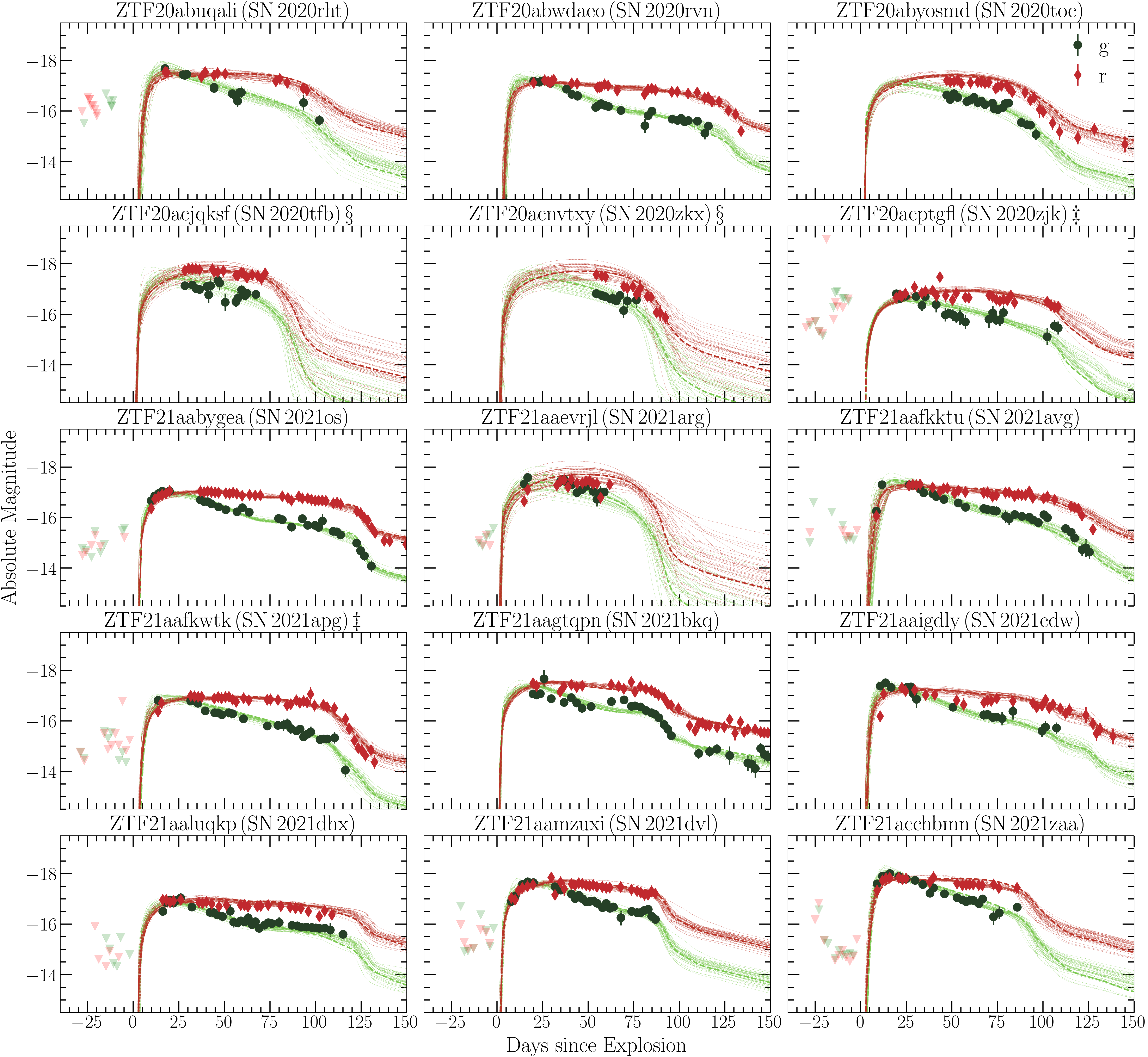}
    \caption{Continued from Figure \ref{fig:grid}. }
    \label{fig:grid_3}
\end{figure*}

\begin{figure*}[ht!]
	\centering
	\includegraphics[width=\textwidth]{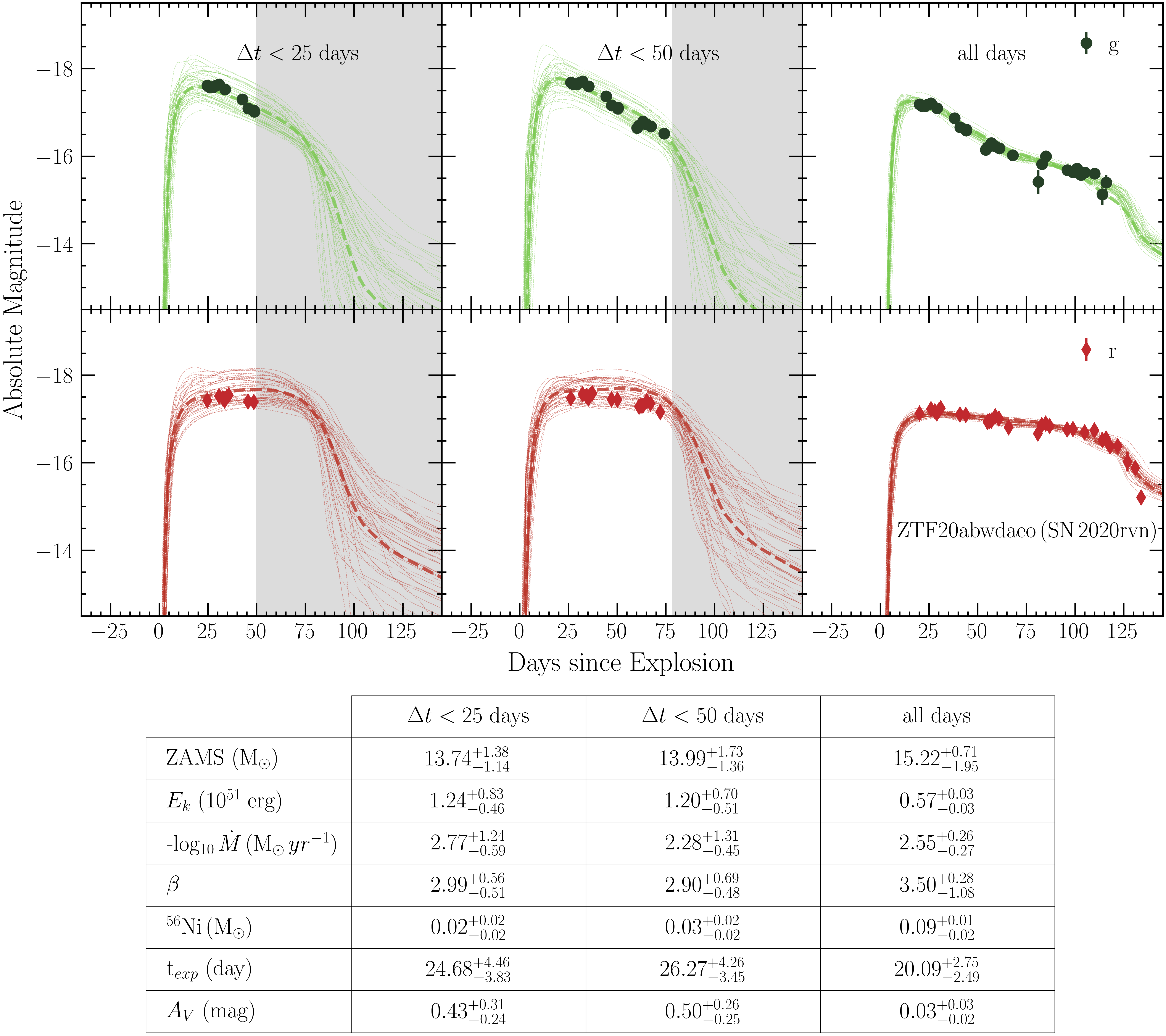}
    \caption{Continued from Figure \ref{fig:dkrm} for event ZTF20abwdaeo (SN 2020rvn).}
    \label{fig:daeo}
\end{figure*}

\begin{figure*}[ht!]
	\centering
	\includegraphics[width=\textwidth]{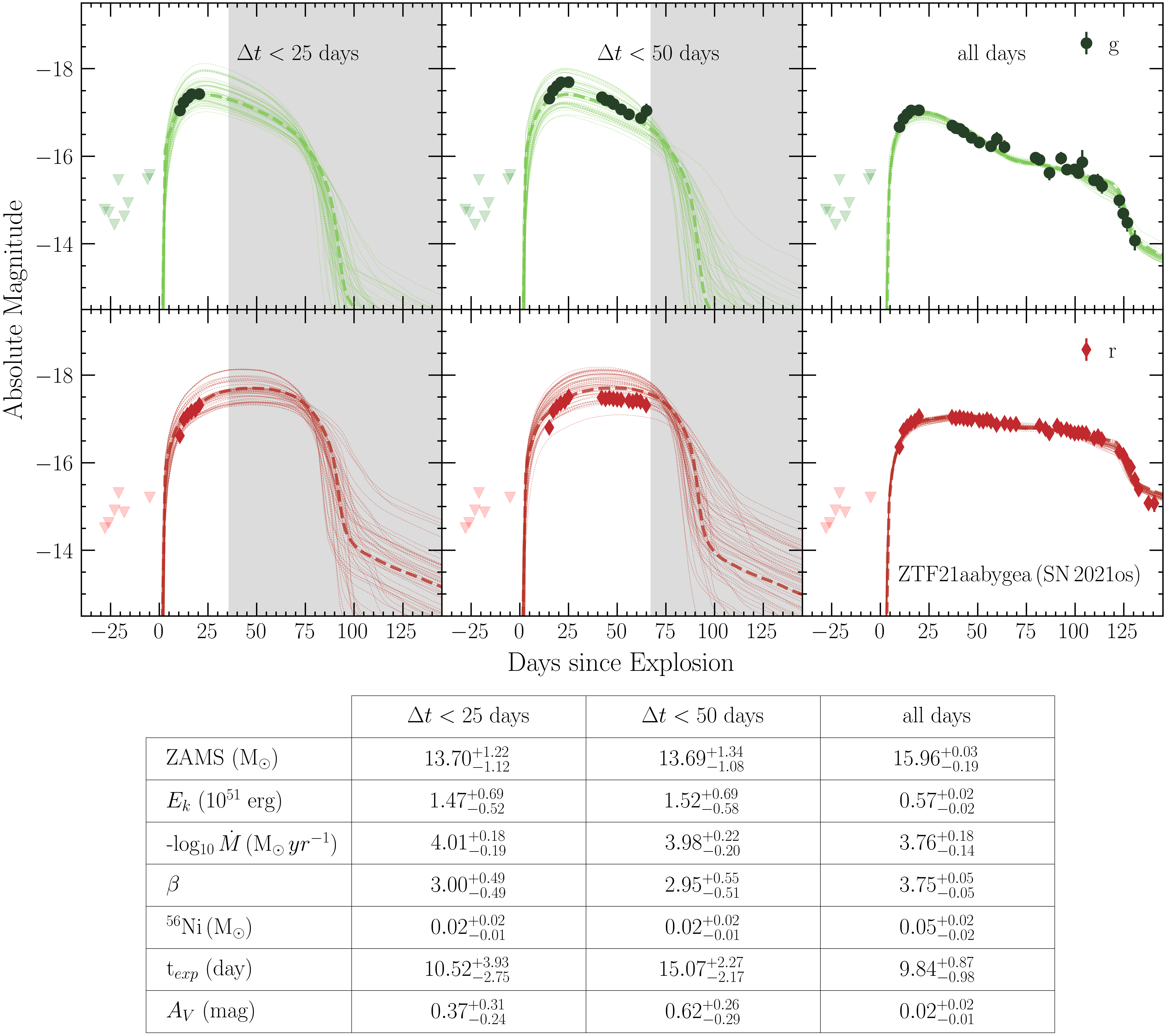}
    \caption{Continued from Figure \ref{fig:dkrm} for event ZTF21aabygea (SN 2021os).}
    \label{fig:multi_epoch_gyea}
\end{figure*}

\begin{figure*}[ht]
	\centering
	\includegraphics[width=\textwidth]{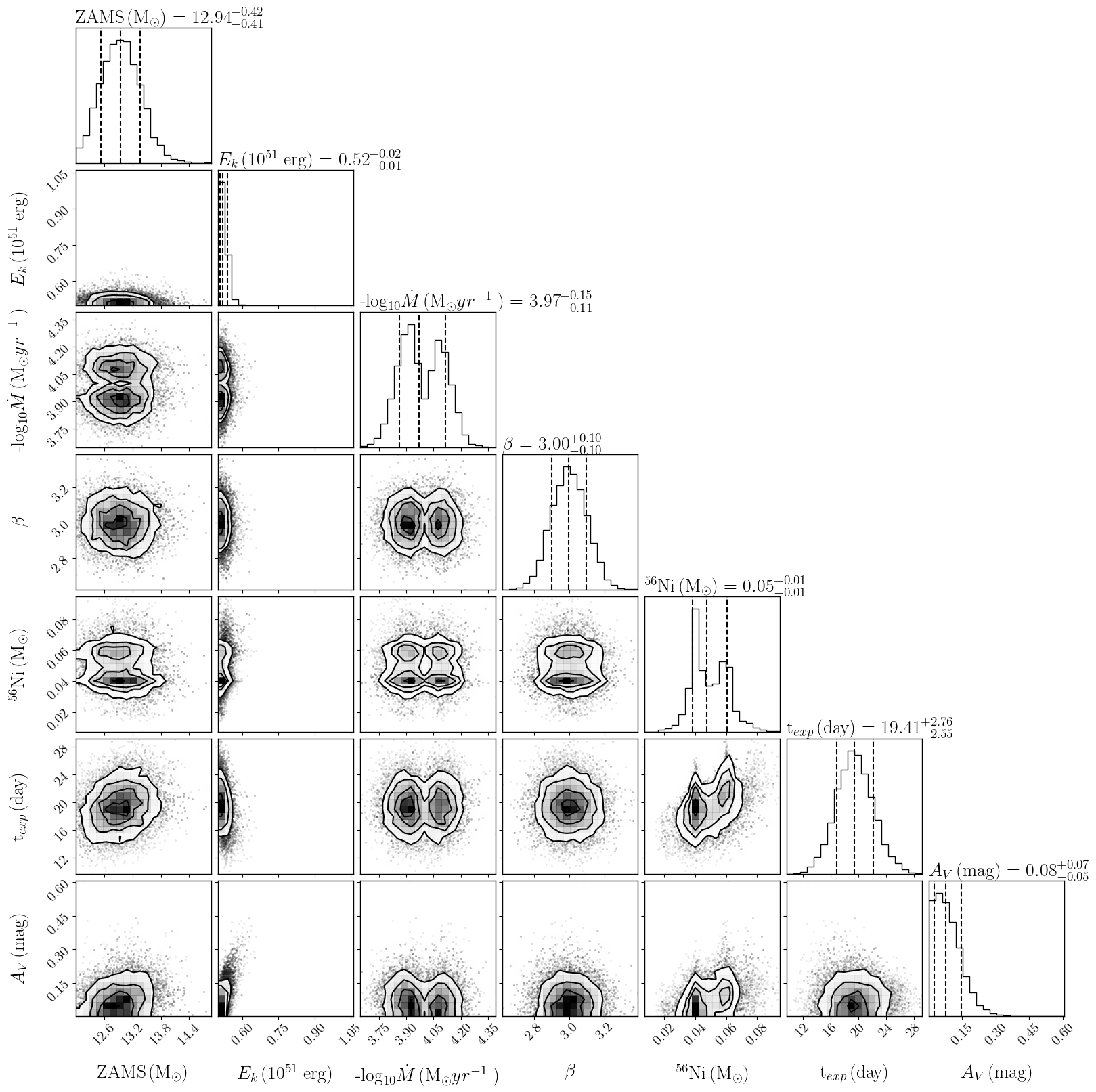}
    \caption{Corner plot showing the posterior probability distribution of various parameters for the event ZTF20acptgfl (SN 2020zjk).}
    \label{fig:corner}
\end{figure*}
\clearpage
\begin{figure*}[ht]
	\centering
	\includegraphics[width=\textwidth]{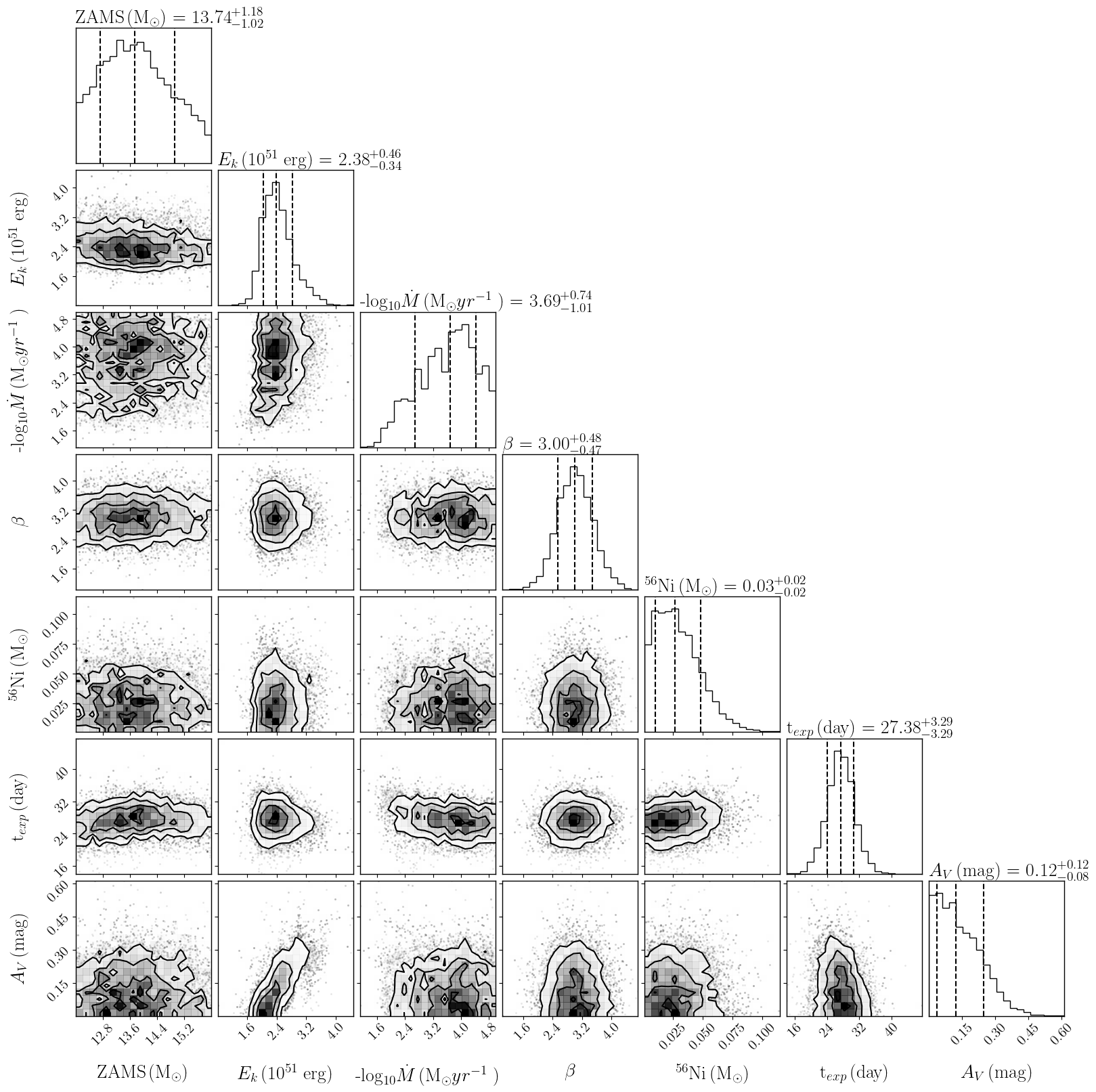}
    \caption{Corner plot showing the posterior probability distribution of various parameters for the event ZTF20aausahr (SN 2020hgm).}
    \label{fig:corner2}
\end{figure*}
\clearpage
\begin{figure*}[ht]
	\centering
	\includegraphics[width=\textwidth]{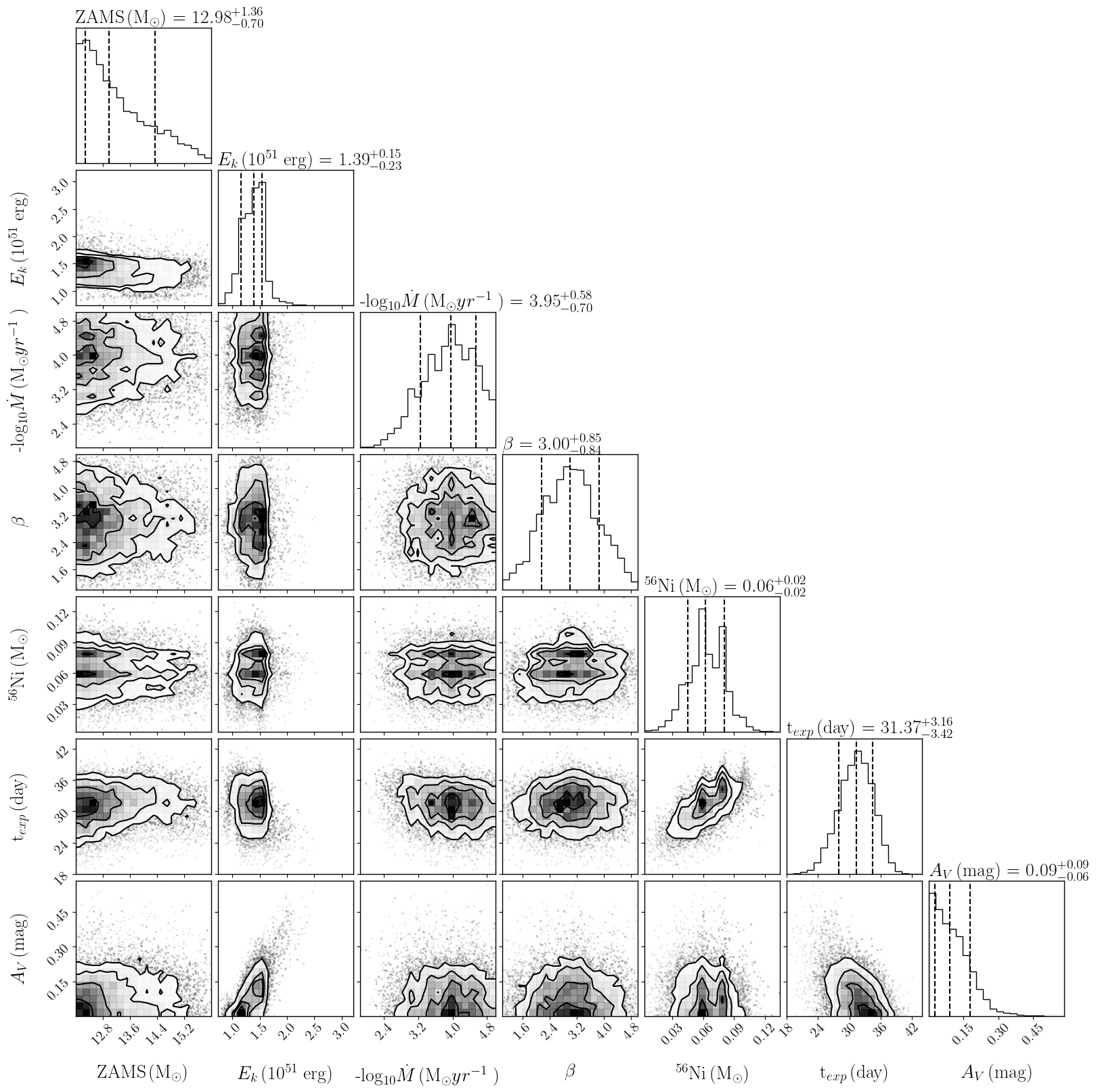}
    \caption{Corner plot showing the posterior probability distribution of various parameters for the event ZTF19abqyouo (SN 2019pbk).}
    \label{fig:corner3}
\end{figure*}

\section{Multi-epoch Evolution of Posterior Distribution of Parameters}

We performed a kernel density estimation (KDE) analysis in order to find the modality of the posterior distributions at various epochs. The samples in the posterior distribution were collected and smoothed using Silverman's bandwidth with a gaussian kernel. The KDE approximated distribution was then used to calculate the number of modes at every epoch. The modes were found by identifying inflection points in the distribution i.e positions where the first derivative changes the sign. 

\begin{figure}
\gridline{\fig{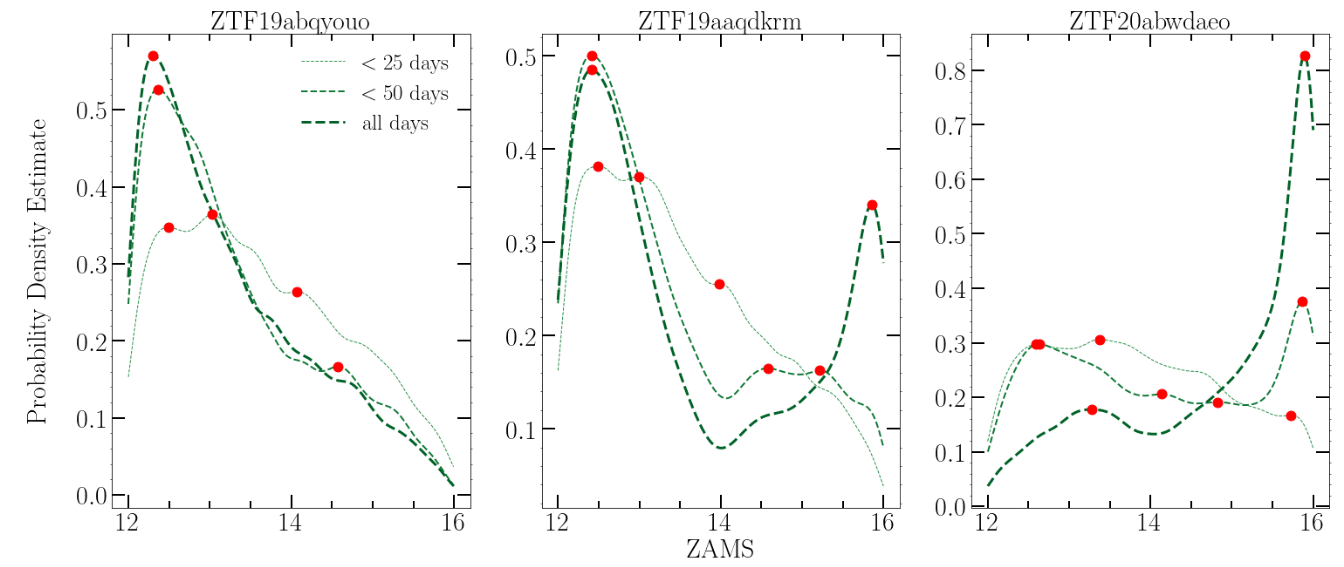}{0.9\textwidth}{(a)}}
\gridline{\fig{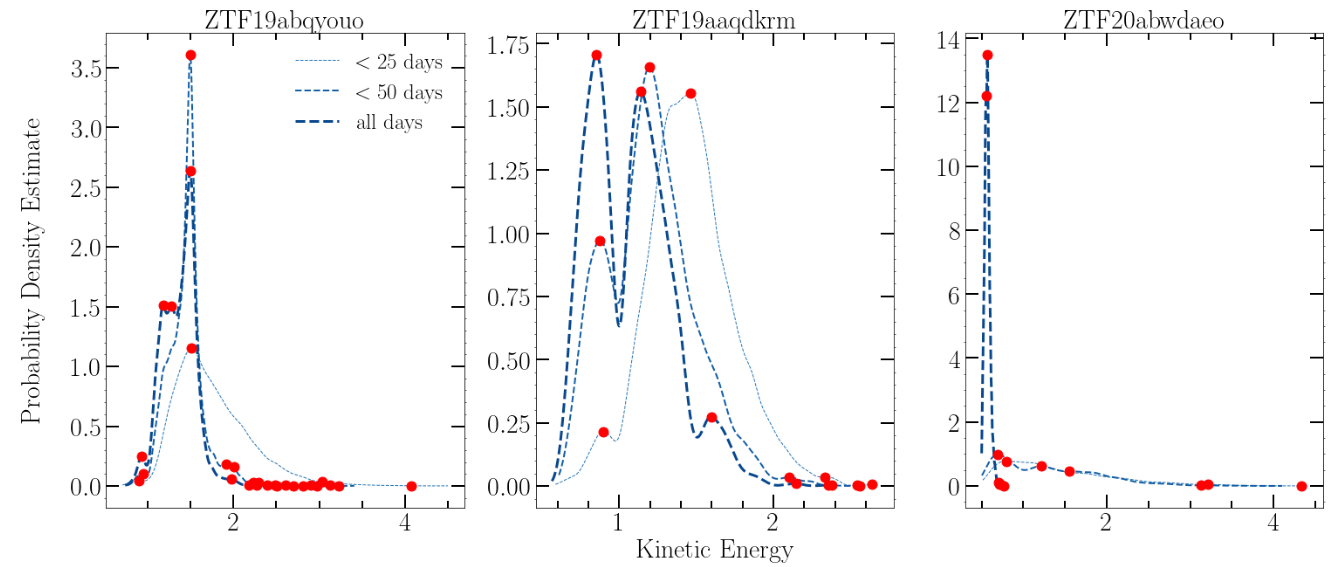}{0.9\textwidth}{(b)}}
\gridline{\fig{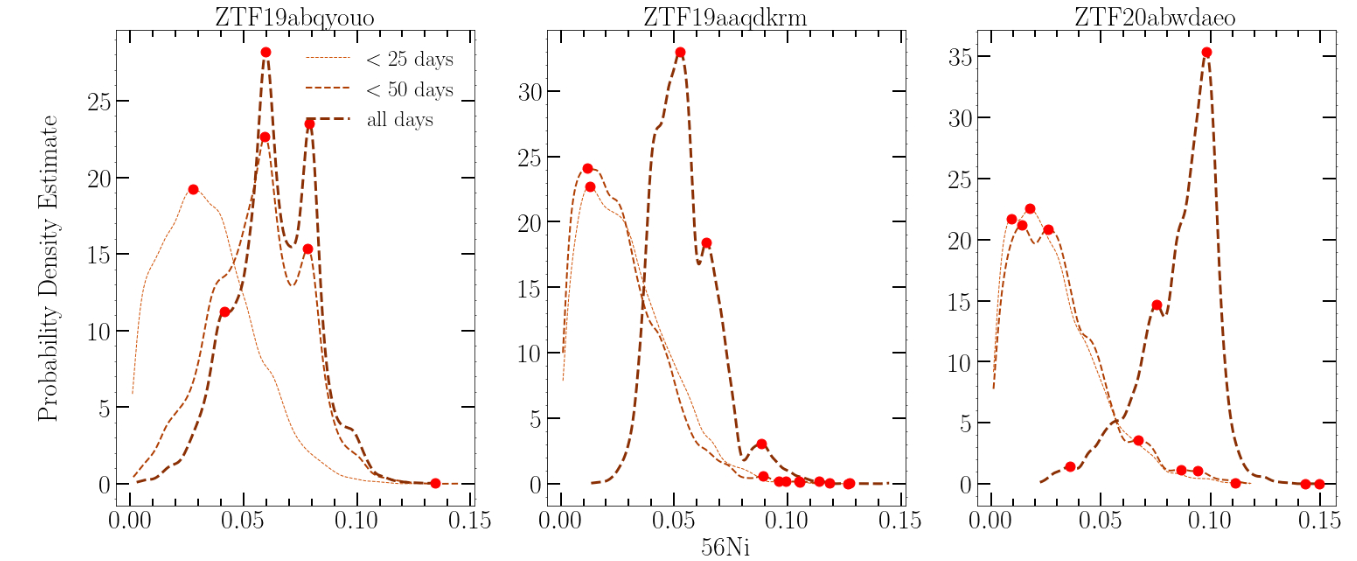}{0.9\textwidth}{(c)}}
\caption{Kernel Density Estimates (KDE) of physical parameters along with modes represented by red circles obtained at three epochs for three ZTF events. The order of physical parameters from top to bottom row are as follows: (a) ZAMS mass, (b) Kinetic Energy and (c) $^{56}$Ni mass.}
\label{fig:kde}
\end{figure}

We found that all the objects have multi-modal posteriors for at least one parameter in our analysis. From this analysis, we conclude that the change in the posteriors for physical parameters over different epochs cannot be generalized for all the events. Figure \ref{fig:kde} shows examples of multi-modal posteriors for ZAMS, Kinetic Energy and $^{56}$Ni for three ZTF events The red circles represent different modes found in the distribution using inflection point analysis. We note that the modes of the distribution for kinetic energy shift from higher to lower values as time proceeds in Figure \ref{fig:kde} for each event as discussed in the paper. The trend in $^{56}$Ni with time is reflected in the modes with earlier epochs favouring lower values as compared with final epochs. The degeneracies in parameter space as discussed in \ref{subsec:degeneracies} are clearly reflected in these distributions through multi-modality.

\clearpage

\bibliographystyle{aasjournal.bst}
\bibliography{main_bib}

\end{document}